\documentclass[conference]{IEEEtran}
\IEEEoverridecommandlockouts
\ifCLASSOPTIONcompsoc
\usepackage[nocompress]{cite}
\else
\usepackage{cite}
\fi
\usepackage{amsmath,amssymb,amsfonts}
\usepackage{algorithmic}
\usepackage{graphicx}
\usepackage{textcomp}
\usepackage{listings}
\usepackage{multirow}  
\usepackage{array}
\usepackage{booktabs}  
\usepackage{tabularx}
\usepackage{makecell}
\usepackage{array}
\usepackage{ragged2e}
\usepackage{calc}
\usepackage{adjustbox}
\usepackage{xurl}
\usepackage{tikz}
\usepackage[hidelinks]{hyperref}
\usepackage{float}

\usepackage{xcolor}

\usepackage[most]{tcolorbox}
\tcbuselibrary{breakable, skins}
\usepackage{soul}
\usepackage{soulutf8}

\definecolor{respBG}{HTML}{D6E5F5}\definecolor{respFG}{HTML}{042C53}
\definecolor{cogBG}{HTML}{DAEBC8}\definecolor{cogFG}{HTML}{173404}
\definecolor{partBG}{HTML}{F5DDD3}\definecolor{partFG}{HTML}{4A1B0C}
\definecolor{learnBG}{HTML}{DEDCFA}\definecolor{learnFG}{HTML}{26215C}
\definecolor{aiBG}{HTML}{D0EBE0}\definecolor{aiFG}{HTML}{04342C}
\definecolor{contBG}{HTML}{F5DFB8}\definecolor{contFG}{HTML}{412402}
\definecolor{enforceFG}{HTML}{222222}
\definecolor{enforceWrapBG}{HTML}{ECECEC}

\newtcbox{\pillshort}[2][respBG]{%
  on line, colback=#1, colframe=#1, coltext=#2,
  boxrule=0.5pt, arc=3pt, outer arc=3pt, boxsep=0pt,
  left=5pt,right=5pt,top=2.3pt,bottom=2.3pt,
  fontupper=\bfseries\footnotesize,
  nobeforeafter, tcbox raise base,
}

\newtcbox{\eboxsolid}{%
  on line, colback=white, colframe=enforceFG, coltext=enforceFG,
  boxrule=0.5pt, arc=3pt, outer arc=3pt, boxsep=0pt,
  left=5pt,right=5pt,top=2.3pt,bottom=2.3pt,
  fontupper=\bfseries\footnotesize,
  nobeforeafter, tcbox raise base,
}

\newtcbox{\eboxdashed}{%
  enhanced, on line,
  colback=white, colframe=white, coltext=enforceFG,
  boxrule=0pt, arc=3pt, outer arc=3pt, boxsep=0pt,
  left=5pt,right=5pt,top=2.3pt,bottom=2.3pt,
  fontupper=\bfseries\footnotesize,
  nobeforeafter, tcbox raise base,
  borderline={0.6pt}{0pt}{enforceFG, dash pattern={on 1.6pt off 1.2pt}},
}

\newtcbox{\eboxdotted}{%
  enhanced, on line,
  colback=white, colframe=white, coltext=enforceFG,
  boxrule=0pt, arc=3pt, outer arc=3pt, boxsep=0pt,
  left=6pt,right=6pt,top=2.8pt,bottom=2.8pt,
  fontupper=\bfseries\footnotesize,
  nobeforeafter, tcbox raise base,
  borderline={0.9pt}{0pt}{enforceFG, line cap=round, dash pattern={on 0pt off 2pt}},
}

\newtcbox{\eboxdouble}{%
  enhanced, on line,
  colback=white, colframe=enforceFG, coltext=enforceFG,
  boxrule=0.4pt, arc=3pt, outer arc=3pt, boxsep=0pt,
  left=5.5pt,right=5.5pt,top=2.4pt,bottom=2.4pt,
  fontupper=\bfseries\footnotesize,
  nobeforeafter, tcbox raise base,
  borderline={0.4pt}{1.2pt}{enforceFG},
}

\newtcbox{\eboxthick}{%
  on line, colback=white, colframe=enforceFG, coltext=enforceFG,
  boxrule=1.2pt, arc=3pt, outer arc=3pt, boxsep=0pt,
  left=5pt,right=5pt,top=2.3pt,bottom=2.3pt,
  fontupper=\bfseries\footnotesize,
  nobeforeafter, tcbox raise base,
}

\newcommand{\Pone}{\pillshort[respBG]{respFG}{P1}}
\newcommand{\Ptwo}{\pillshort[respBG]{respFG}{P2}}
\newcommand{\Pthree}{\pillshort[respBG]{respFG}{P3}}
\newcommand{\Pfour}{\pillshort[cogBG]{cogFG}{P4}}
\newcommand{\Pfive}{\pillshort[cogBG]{cogFG}{P5}}
\newcommand{\Psix}{\pillshort[cogBG]{cogFG}{P6}}
\newcommand{\Pseven}{\pillshort[partBG]{partFG}{P7}}
\newcommand{\Peight}{\pillshort[partBG]{partFG}{P8}}
\newcommand{\Pnine}{\pillshort[learnBG]{learnFG}{P9}}
\newcommand{\Pten}{\pillshort[learnBG]{learnFG}{P10}}
\newcommand{\Peleven}{\pillshort[aiBG]{aiFG}{P11}}
\newcommand{\Ptwelve}{\pillshort[contBG]{contFG}{P12}}

\newcommand{\Eone}{\eboxsolid{E1}}
\newcommand{\Etwo}{\eboxsolid{E2}}
\newcommand{\Ethree}{\eboxdashed{E3}}
\newcommand{\Efour}{\eboxdashed{E4}}
\newcommand{\Efive}{\eboxdotted{E5}}
\newcommand{\Esix}{\eboxdotted{E6}}
\newcommand{\Eseven}{\eboxdotted{E7}}
\newcommand{\Eeight}{\eboxdouble{E8}}
\newcommand{\Enine}{\eboxdouble{E9}}
\newcommand{\Eten}{\eboxthick{E10}}
\newcommand{\Eeleven}{\eboxthick{E11}}

\lstdefinestyle{query}{
  basicstyle=\ttfamily\footnotesize,
  breaklines=true,
  breakatwhitespace=false,
  columns=fullflexible,
  keepspaces=true,
  frame=single,
  xleftmargin=0.6em,
  xrightmargin=0.2em
}
\def\BibTeX{{\rm B\kern-.05em{\sc i\kern-.025em b}\kern-.08em
    T\kern-.1667em\lower.7ex\hbox{E}\kern-.125emX}}
\begin{document}

\title{Exploring the Design Space of LLM-Based Programming Support in CS Education: A Scoping Review through the Lens of Assistance Governance}

\author{%
  \IEEEauthorblockN{Minsun Kim}
  \IEEEauthorblockA{%
    \textit{Virginia Tech}\\
    Blacksburg, USA\\
    minsunkim@vt.edu
  }
  \and
  \IEEEauthorblockN{S. Moonwara A. Monisha}
  \IEEEauthorblockA{%
    \textit{Virginia Tech}\\
    Blacksburg, USA\\
    msheikhmoonwaraa@vt.edu
  }
  \and
  \IEEEauthorblockN{Zihan Wu}
  \IEEEauthorblockA{%
    \textit{University of Maine}\\
    Orono, USA\\
    zihan.wu@maine.edu
  }
  \and
  \IEEEauthorblockN{David H. Smith IV}
  \IEEEauthorblockA{%
    \textit{Virginia Tech}\\
    Blacksburg, USA\\
    dhsmith4@vt.edu
  }
}

\maketitle

\begin{abstract}
As large language models (LLMs) become integrated into programming education, learner-facing systems increasingly differ in how that assistance is bounded, enacted, and controlled. These governance decisions are often described implicitly, making it difficult to compare systems in educationally meaningful ways. To address this gap, we conduct a scoping review and qualitative synthesis of 90 peer-reviewed LLM-based programming support systems in CS education. We analyze assistance governance through three dimensions, which we refer to collectively as PEA: Policy, capturing what forms of help are allowed or restricted; Enforcement, capturing how those boundaries are operationalized through interaction and system behavior; and Authority, capturing who can configure, adapt, or override them during use. Our findings show that systems often share similar pedagogical goals, but implement those goals through varied enforcement mechanisms. At the same time, authority remains highly centralized in system logic, with fewer systems giving learners or instructors runtime control. This work contributes PEA as a three-dimensional analytic lens, a governance codebook empirically refined within these dimensions, and a map of underexplored configurations in the current design space of LLM-based programming support. By making these explicit and comparable, PEA offers a vocabulary for analyzing existing systems and designing future tools that are pedagogically bounded, configurable, and accountable.
\end{abstract}


\section{Introduction}
In recent years, large language models (LLMs) have been increasingly adopted in educational tools. In computer science (CS) education, LLMs are especially relevant because they can generate, explain, debug, and revise code across a range of programming tasks \cite{agbo_computing_2025, raihan_large_2025}. However, LLM assistance is unlikely to work as a one-size-fits-all form of support. Prior work shows that instructors adapt AI code generation and explanation tools in relation to course goals, while novice learners use LLM-based code generators in varied ways across programming tasks \cite{lau_ban_2023, kazemitabaar_how_2023}. This suggests that the educational value of LLM assistance depends on how support is bounded and controlled in learning environments.

Prior reviews have mapped uses, tools, outcomes, practices, and risks of GenAI in computing education \cite{agbo_computing_2025, raihan_large_2025}, but they have not typically compared how learner-facing assistance is bounded, operationalized, and controlled. We use the term ``assistance governance'' to describe this design problem: how learning support systems define, implement, and allocate control over learner-facing assistance.

To address this gap, we organize assistance governance around three practical design questions: what forms of assistance are allowed or restricted (Policy), how those boundaries are implemented (Enforcement), and who can control them during use (Authority). This What--How--Who structure was initially informed by the 5W1H
design heuristic; the three dimensions were specified before coding, while their subcategories and decision rules were refined through engagement with the reviewed corpus\cite{jia_5w1h_2016}. We refer to these dimensions as Policy, Enforcement, and Authority (PEA), and use them to analyze
governance-relevant design decisions across 90 peer-reviewed, learner-facing LLM-based programming support systems published
between January 2023 and March 2026. 

In analyzing this corpus, this study addresses the following research questions:

\begin{itemize}
\item \textbf{RQ1:} What policy, enforcement, and authority (PEA) categories characterize assistance governance in LLM-based programming support systems?
\item \textbf{RQ2:} How are these categories distributed and combined across systems, and what underexplored configurations emerge from the resulting design space?
\end{itemize}

This paper makes three contributions. First, we introduce PEA as a governance-focused analytic lens for comparing learner-facing LLM programming support systems. Second, we provide a scoping review and qualitative synthesis of 90 systems, mapping how assistance boundaries, enforcement mechanisms, and authority arrangements are represented across the literature. Third, we identify recurring governance patterns and design gaps, including reliance on prompt-based constraints, centralized authority, and underexplored PEA configurations.

\section{Related Work}
\subsection{Prior Reviews of LLMs in CS Education}

Recent review papers show that generative AI has become a major topic in computing education. Broad reviews have mapped its use across programming tasks, teaching practices, classroom settings, and educator responses. They also discuss recurring concerns, including over-reliance, bias, misinformation, and academic misconduct \cite{agbo_computing_2025, prather_beyond_2025, raihan_large_2025, pirzado_navigating_2024}. For example, Agbo et al. and Raihan et al. synthesize tools, uses, and risks of GenAI in computing education, but their review categories do not distinguish whether a system restricts solution-level help, how that restriction is enforced, or whether learners or instructors can modify it during use \cite{agbo_computing_2025, raihan_large_2025}.

More focused reviews narrow this picture. Qiao et al.\ examine GenAI assistants for code comprehension and argue that this area has not been examined closely in broader reviews, especially in terms of tools, approaches, and evaluation methods \cite{qiao_systematic_2026}. These reviews provide a strong account of where and why LLMs are being used in CS education, but they offer less support for comparing how assistance is bounded, enacted, and controlled across learner-facing systems.

\subsection{Why Assistance Governance Requires a Design-Space Account}

LLM-based programming support systems do not differ only in what models they use or what tasks they support. They also differ in how assistance is organized: whether systems provide direct solutions, restrict help to hints or explanations, require learner effort before feedback, or allow assistance to vary by task or instructional context. These choices can substantially change the pedagogical role of a system, yet they are not well captured by existing labels.

Prior work motivates these questions but does not provide a shared vocabulary for comparing them across systems. Studies of novice programmers show that learners use LLM-based code generators through varied approaches, including single-prompt solution generation, step-by-step generation, hybrid use, and manual coding, raising concerns about over-reliance and self-regulation \cite{kazemitabaar_how_2023, kazemitabaar_codeaid_2024, liu_tool_2026}. Work on cognitive engagement techniques further shows that the same underlying capability—AI-generated code—can be shaped through different interaction designs to encourage deeper learner engagement \cite{kazemitabaar_exploring_2025}. These studies motivate the policy question: what forms of assistance should be available, delayed, or restricted in learning contexts?

Other work motivates the authority question. Participatory design research on AI-supported programming tools shows that learners and instructors differ in their preferences for autonomy, guidance, and adaptive control \cite{wu_learner_2025}. Learners may prefer greater autonomy, while instructors may prefer stronger system guidance to prevent cognitive overload or misuse. These findings suggest that assistance governance is not only about what a system provides, but also about who can configure or adjust assistance boundaries during use.

Recent design-space work on LLM-based AI coding assistants provides a precedent for comparing systems through recurring design dimensions \cite{lau_design_2025}. However, that work focuses broadly on AI coding assistants across academic and industry systems. Our review narrows this design-space perspective to learner-facing programming support in CS education.

\subsection{Positioning PEA as an Analytic Lens}

PEA began with a 5W1H-inspired heuristic \cite{jia_5w1h_2016}, which uses interrogative dimensions---Who, What, When, Where, Why, and How---to structure problem analysis. Rather than adopting all six, we selectively used its What--How--Who structure as a conceptual starting point: \textit{What} maps to Policy (what assistance is permitted or restricted), \textit{How} maps to Enforcement (how those boundaries are operationalized), and \textit{Who} maps to Authority (who can configure or override them during use). The remaining three dimensions are absorbed by our scope and inclusion criteria: \textit{Where} and \textit{When} are fixed by our focus on learner-facing systems used during programming tasks in CS education, and \textit{Why} is addressed by the shared goal of supporting CS education.

We subsequently grounded these dimensions in relevant research. Policy draws on work on scaffolding, help seeking, over-reliance, and academic integrity \cite{reiser_scaffolding_2004, aleven_help_2016, prather_beyond_2025, pirzado_navigating_2024}. Enforcement draws on access-control distinctions and Human--AI interaction guidelines \cite{sandhu_role_1996, amershi_guidelines_2019}. Authority draws on human--automation control allocation, learner agency, and classroom orchestration \cite{parasuraman_model_2000, wu_learner_2025, dillenbourg_design_2013}.

These traditions clarify the meaning of each dimension but do not directly provide a vocabulary for comparing assistance governance across learner-facing LLM programming support systems. The three PEA dimensions were specified before corpus coding, while subcategories and decision rules were refined through iterative engagement with the reviewed literature. We therefore use PEA as a corpus-grounded analytic lens rather than a complete or externally validated theory of governance.

\section{Methods}
We adopted a hybrid approach that combines a scoping review with a qualitative synthesis. The scoping review was used to identify, screen, and assemble a transparent and reproducible corpus of relevant studies. The qualitative synthesis was then used to analyze governance-relevant design decisions across the included systems and to populate a design space organized by the PEA dimensions  \cite{barnett-page_methods_2009}. The three PEA dimensions were specified before corpus coding as an analytic lens, while the subcategories within each dimension were refined through iterative qualitative coding and synthesis. Policy characterizes what forms of assistance are permitted. Enforcement characterizes how assistance boundaries are implemented through interaction and system behavior. Authority characterizes how decision rights over these boundaries are allocated in the system. We use the term system broadly to include research prototypes and deployed tools studied in educational contexts, provided that the paper reported governance-relevant conditions shaping how assistance was bounded, enacted, or controlled in practice.

\begin{figure}
    \centering
    \includegraphics[width=\linewidth]{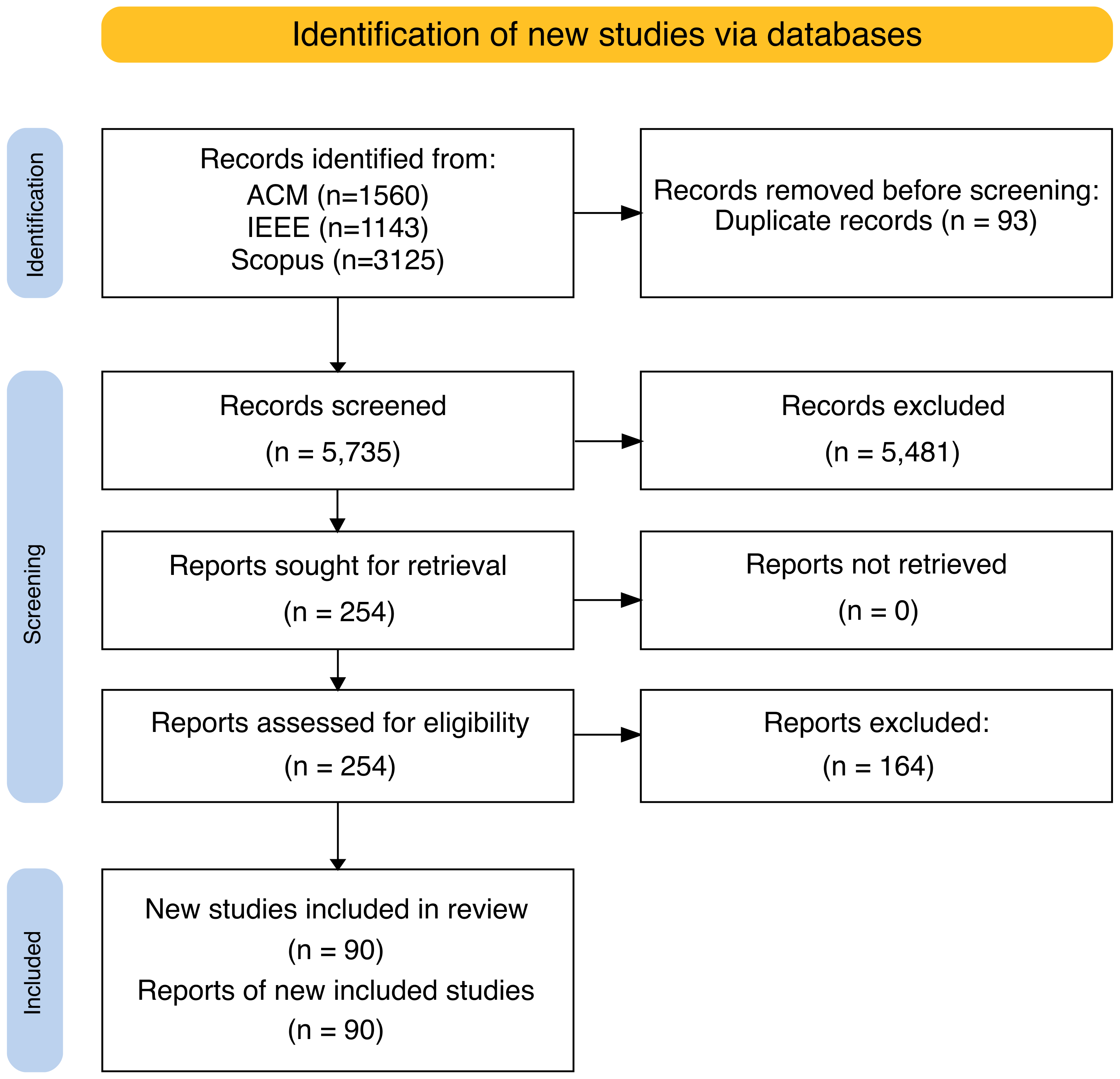}
    \caption{PRISMA\cite{page_prisma_2021} flow diagram of the study selection process. The diagram summarizes identification, screening, eligibility assessment, and final inclusion of papers reviewed in this study.}
    \label{fig:prisma}
\end{figure}

\subsection{Search Strategy, Sources, and Study Selection}

We conducted a structured search of peer-reviewed literature in ACM Digital Library, IEEE Xplore, and Scopus. Searches were conducted on March 10, 2026 and covered records published from January 2023 through the search date. We set January 2023 as the starting point because ChatGPT was launched on November 30, 2022, shortly before the rapid growth of LLM-based educational tool research. The search combined four concept groups: LLM-related terms, CS/programming education terms, learner-facing system/tool terms, and tool design or development terms. A representative compressed search string was:

\begin{quote}
\small
(("large language model*" OR LLM* OR "generative AI" OR ChatGPT OR GPT OR Gemini OR Claude OR DeepSeek) AND
("computer science education" OR "computing education" OR "programming education" OR "introductory programming" OR "novice programmer*" OR student* OR instructor* OR course* OR classroom*) AND
(chatbot* OR tool* OR system* OR assistant* OR platform* OR tutor* OR tutoring) AND
(design* OR creat* OR introduc* OR develop* OR prototype* OR modif*))
\end{quote}

The string was adapted to the field syntax and filtering options of each database, with searches applied to title and abstract fields where supported. The full keyword-based filtering criteria used for secondary screening are provided in the supplemental material (Appendix~\ref{app:filtering-criteria},~\ref{app:search-query}).

The screening flow and outcomes are reported using a PRISMA-ScR flow diagram (Fig.~\ref{fig:prisma}) \cite{page_prisma_2021}. All retrieved records were initially screened by a single reviewer at the title and abstract level. A second reviewer independently screened a randomly selected subset of 1,010 papers, with disagreements resolved through discussion and consensus \cite{chinh_ways_2019}. Following this stage, 254 papers were selected for full-text screening. Three reviewers participated in full-text screening: one reviewer screened all papers, while the other two each screened a randomly assigned half after a 20-paper calibration exercise.

We included papers that described or evaluated LLM-based programming assistance in a computer science learning context, where programming learning was an intended outcome. Eligible papers had to report a learner-facing tool, system, or prototype designed for computer science or programming education, include at least one meaningful adaptation beyond direct use of a general-purpose LLM interface or unmodified API, and provide sufficient detail to support governance extraction. We excluded papers focused primarily on course design, curriculum design, instructional planning, learning analytics, literature reviews, benchmark studies, survey studies, perception- or interview-only studies, direct use of ChatGPT or an unmodified API, papers under five pages, and papers without a user study. Details in Appendix~\ref{app:screening-audit}.

\subsection{Descriptive Analysis}
\label{sub:descriptive}

\begin{table}[t]
\caption{Descriptive Analysis Framework}
\label{tab:descriptive_framework}
\centering
\footnotesize
\begin{tabularx}{\columnwidth}{l l X}
\toprule
\textbf{\makecell[l]{Dimension}} & \textbf{Category} & \textbf{Description} \\
\midrule

\makecell[l]{Target\\Learner}
& Introductory 
& Beginner-level learners in introductory computing contexts \\
& Not introductory 
& Learners beyond the introductory level \\
& Unspecified
& Learner level not reported \\

\midrule
Supervision 
& Supervised 
& Controlled setting; outside resources restricted \\
& Semi-supervised 
& Some oversight; outside resources limited \\
& Unsupervised 
& No direct oversight; outside resources allowed \\
& Unspecified
& Supervision level not reported \\

\midrule
\makecell[l]{Study\\Environment}
& In-class 
& Course-linked setting (e.g., lecture, lab, discussion) \\
& External 
& Researcher-organized setting outside regular class \\
& Home/Field 
& Open or naturalistic use outside direct control \\
& Unspecified 
& Environment not reported \\

\midrule
\makecell[l]{Study\\Length}
& One semester 
& About one academic semester \\
& Short-term 
& Less than one semester \\
& Longitudinal 
& More than one semester \\
& Unspecified 
& Length not reported \\

\midrule
\makecell[l]{Methodology\\(Quant.)}
& Experimental 
& Randomized experiment \\
& Quasi-experimental 
& Non-randomized intervention or comparison \\
& Survey 
& Questionnaire-based quantitative study \\
& Secondary data 
& Analysis of logs or existing records \\
& Correlational 
& Analysis of variable relationships \\
& None
& No quantitative analysis \\

\midrule
\makecell[l]{Methodology\\(Qual.)}
& Case study 
& In-depth bounded case analysis \\
& Ethnography/Obs. 
& Naturalistic observation \\
& Interview 
& Individual interview-based study \\
& Focus group 
& Group discussion-based study \\
& Document analysis 
& Analysis of documents or artifacts \\
& Thematic analysis 
& Thematic coding without explicit design \\
& None
& No Qualitative Analysis \\

\bottomrule
\end{tabularx}
\end{table}

We coded descriptive characteristics of the corpus, including target learners, supervision level, study environment, study duration, and research methodology (Table~\ref{tab:descriptive_framework}).

\subsection{PEA Governance Extraction and Codebook Development}

We employed an LLM\footnote{OpenAI ChatGPT, accessed through a custom GPT configured with a structured coding prompt. Details in Appendix~\ref{app:llm-agent}.} as a human-in-the-loop analytic aid during the preliminary coding phase. The LLM was used only to generate provisional evidence candidates and possible policy–enforcement annotations. It was not treated as a coder or rater, and no LLM-generated label was accepted without human review \cite{tai_examination_2024, blondeel_practical_2025}. Because LLM outputs may contain misclassifications or unsupported inferences, all suggestions were treated as provisional and manually reviewed \cite{tai_examination_2024}.

After a 20-paper pilot phase, the full corpus was processed with the LLM to generate candidate policy--enforcement pairs across all papers. The research team reviewed these outputs to consolidate overlapping categories, clarify category boundaries, and refine code definitions and decision rules. This process resulted in a finalized codebook of subcategories under the pre-specified PEA dimensions. Thus, the PEA dimensions organized the analysis deductively, whereas the subcategories and decision rules were refined iteratively from the corpus. Authority was coded separately from policy and enforcement by assigning each paper to one of four categories: system only, system + learner, system + instructor, or system + learner + instructor. For Authority, we coded learner or instructor control only when the paper reported explicit runtime system features for adjusting assistance boundaries, such as selecting hint levels, enabling or disabling solution access, configuring feedback types, changing scaffolding settings, or overriding system restrictions.

Inter-rater reliability was assessed only on human coding decisions made using the finalized codebook. We did not compute LLM-human reliability because the LLM was not positioned as an independent coder, it was used only to surface provisional evidence candidates and possible labels for human review. Two human reviewers independently coded the same subset of 20 papers based on textual evidence from the full papers. Because Policy and Enforcement were coded as multi-label dimensions, each label was treated as a binary present/absent decision, and unweighted Cohen's $\kappa$ was calculated separately for each label. We report macro-averaged $\kappa$ across labels. Agreement was substantial for both Policy (macro-averaged Cohen's $\kappa = .705$) and Enforcement (macro-averaged Cohen's $\kappa = .679$), following commonly used interpretation guidelines in which $\kappa$ values between .61 and .80 indicate substantial agreement \cite{landis_measurement_1977}. Disagreements were resolved through consensus discussion \cite{chinh_ways_2019}.

The per-paper PEA coding for all 90 included studies is provided in the supplemental material (Appendix~\ref{app:paper-level-coding}).

\section{Corpus Overview}
\subsection{Descriptive Analysis}
\begin{figure*}[t]
    \centering
    \includegraphics[width=\textwidth]{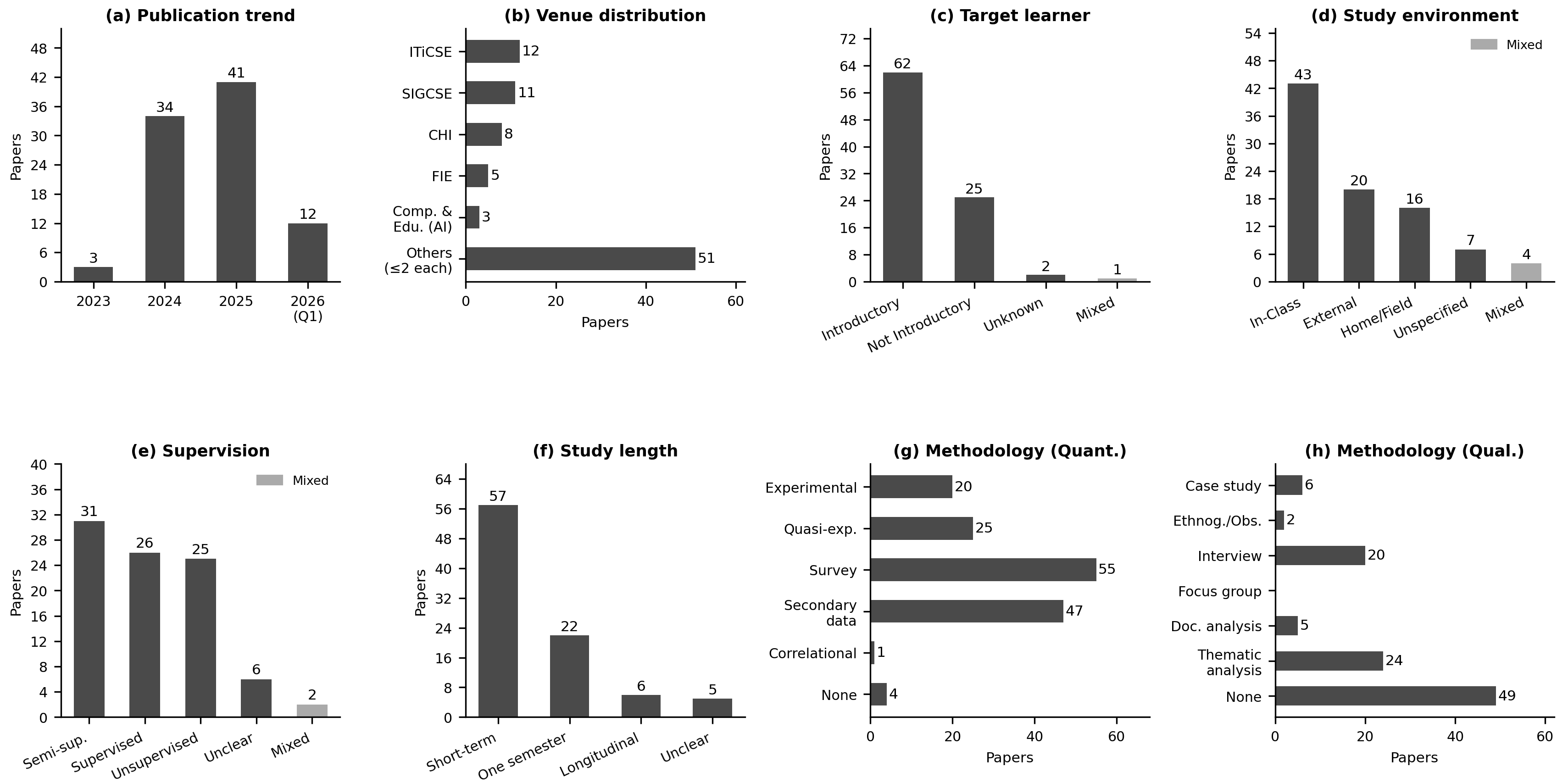}
    \vspace{-0.5cm}
    \caption{Corpus overview of the reviewed studies. The combined figure summarizes (a) publication trend, (b) venue distribution, (c) target learner, (d) study environment, (e) supervision, (f) study length, (g) quantitative methodology, and (h) qualitative methodology.}
    \label{fig:corpus_overview}
    \vspace{-0.5cm}
\end{figure*}
Fig.~\ref{fig:corpus_overview} summarizes the descriptive characteristics of the 90 included studies. The corpus reflects the rapid growth of LLM-based educational tool research following the broad availability of general-purpose models. Publication venues were distributed across CS education (e.g., ITiCSE, SIGCSE), HCI (e.g., CHI), and software engineering communities.

\subsection{Learners and Learning Contexts}
The reviewed systems mostly targeted introductory programming learners (62 studies, 69\%), while 25 studies (28\%) addressed non-introductory contexts and one study (1\%) spanned both introductory and non-introductory learners. In terms of deployment, in-class research settings were most common (43 studies, 48\%), followed by external researcher-managed settings (20 studies, 22\%) and home or field contexts (16 studies, 18\%). Four studies were coded as mixed across two deployment environments. This breadth suggests that LLM-based programming support is being explored across structurally different educational arrangements rather than within a single canonical setting.

\subsection{Study Design and Methodology}
Most studies were short-term (57 studies, 63\%), relying on one-time or brief evaluations, while semester-long deployments accounted for 22 studies (24\%) and longitudinal work beyond a single semester remained rare (6 studies, 7\%). Methodologically, survey-based and secondary data approaches dominated, and over half of the corpus incorporated no qualitative component (49 studies, 54\%), limiting insight into learner experience and process.

\subsection{Associations Across Study Characteristics}
Beyond the individual distributions, several meaningful associations emerged across study characteristics. Supervision type and study environment were strongly associated ($\chi^2(9) = 110.40$, $p < .001$, Cramer's $V = .66$). Supervised studies were predominantly conducted in external researcher-managed settings (17 of 26 studies, 65\%), semi-supervised studies were concentrated in in-class settings (27 of 30 studies, 90\%), and unsupervised studies clustered in home or field contexts (14 of 23 studies, 61\%). This pattern suggests that supervision level is less a freely chosen methodological variable than a structural consequence of deployment context: once a system is used outside an institutional or researcher-controlled setting, maintaining oversight becomes inherently difficult.

Study environment was also associated with study duration ($\chi^2(9) = 29.78$, $p < .001$, Cramer's $V = .34$). All external studies were short-term (20 of 20 studies, 100\%), whereas in-class studies more frequently spanned a semester (17 of 43 studies, 40\%). Home or field deployments accounted for most longitudinal cases in the analyzed subset (4 of 6 studies, 67\%). These patterns suggest that evaluation duration was shaped by deployment context, with external studies tending toward short-term designs and home or field deployments enabling longer-term observation.

Finally, descriptive trends suggest a gradual broadening of methodological practice. Qualitative methods appeared in 23 of 41 studies from 2025 (56\%), compared with 13 of 34 studies from 2024 (38\%). True experimental designs also increased from 5 of 34 studies in 2024 (15\%) to 11 of 41 studies in 2025 (27\%). Across all 90 studies, 41 (45.6\%) incorporated at least one qualitative method, pointing to a potential shift toward more mixed-method and experimentally grounded evaluations in recent work.

\section{FINDINGS: Mapping Assistance Governance Across PEA Dimensions}
\subsection{Policy}
\begin{table*}[!t]
\centering
\footnotesize
\renewcommand{\arraystretch}{1.5}
\setlength{\tabcolsep}{6pt}

\caption{Policy categories}
\label{tab:policy-definitions}

\begin{tabular}{>{\centering\arraybackslash}m{0.07\textwidth} >{\centering\arraybackslash}m{0.026\textwidth} m{0.12\textwidth} m{0.37\textwidth} m{0.30\textwidth}}
\hline
\textbf{Theme} & \textbf{ID} & \textbf{Policy} & \textbf{Definition} & \textbf{Representative Quote} \\ \hline

\multirow{3}{*}{\makecell{\textbf{Response} \\ \textbf{Constraint}}}
& \Pone
& Scope Restriction
& Limits AI assistance to a defined course, topic, assignment, task, role, or instructional context.
& ``A description of the role of the system as an AI tutor. Learning objectives and course-specific information are provided'' \cite{tran_pacing_2026} \\ \cline{2-5}

& \Ptwo
& Direct-Solution \newline Prohibition
& Restricts AI from providing complete answers, full solutions, model responses, or directly usable code.
& ``designed to guide learners toward solutions rather than offer them outright'' \cite{liu_teaching_2024} \\ \cline{2-5}

& \Pthree
& Scaffolding-\newline Oriented Support
& Requires AI assistance to be delivered as hints, next steps, prompts, or incremental guidance rather than final answers.
& ``steering the LLM to cover 3 or 4 issues in a single hint could lead to learner progress more often than honing in on one bug'' \cite{qi_knowledge-component-based_2025} \\ \hline

\multirow{3}{*}{\makecell{\textbf{Cognitive} \\ \textbf{Scaffolding}}}
& \Pfour
& Independent \newline Problem Solving
& Requires or evaluates learners' ability to complete problem-solving steps independently, without relying on AI to perform the task for them.
& ``feedback into the reflective learning process after learners have attempted to solve problems independently'' \cite{chen_coding_2024} \\ \cline{2-5}

& \Pfive
& Pedagogical \newline Alignment
& Requires AI assistance to remain consistent with course objectives, instructional level, lesson structure, or intended learning outcomes.
& ``designed to be a ``pedagogically-minded subject-matter expert'' '' \cite{yang_debugging_2024} \\ \cline{2-5}

& \Psix
& Demonstrating \newline Understanding
& Requires learners to explain, justify, reflect on, or otherwise demonstrate their understanding before, during, or after receiving AI support.
& ``implementing the Traffic Light Methodology, which promotes the structured analysis of algorithmic problems'' \cite{martinez_enhancing_2025} \\ \hline

\multirow{2}{*}{\makecell{\textbf{Participation} \\ \textbf{Support}}}
& \Pseven
& Collaborative \newline Learning Support
& Structures AI assistance to support peer interaction, group work, shared problem solving, or collaborative reflection.
& ``provides feedback from the perspectives of different roles, ensuring a diverse and realistic collaboration experience'' \cite{wang_devcoach_2025} \\ \cline{2-5}

& \Peight
& Reward-Based \newline Engagement
& Motivates learners through rewards, points, badges, rankings, progress indicators, gamified feedback, or other visible achievement structures.
& ``the learners also saw a greyed out trophy icon being colored as yellow.'' \cite{logacheva_evaluating_2024} \\ \hline

\multirow{2}{*}{\makecell{\textbf{Learner} \\ \textbf{Adaptation}}}
& \Pnine
& Learning-Based \newline Personalization
& Adapts AI assistance based on learners' knowledge level, progress, errors, performance, or demonstrated difficulties.
& ``Personalized learning suggestions: Analyzes mistakes and suggests targeted exercises'' \cite{zhao_exploration_2025} \\ \cline{2-5}

& \Pten
& Engagement-Based \newline Personalization
& Adapts AI assistance based on learners' interests, preferences, affective state, motivation, or engagement patterns.
& ``continuously updates dynamic learner personas using behavioral signals, linguistic cues, and meta-dialogue patterns'' \cite{troussas_personagpt_2025} \\ \hline

\makecell{\textbf{AI} \\ \textbf{Literacy}}
& \Peleven
& Responsible AI \newline Use Instruction
& Uses AI assistance to teach learners how to use AI effectively, critically, ethically, or reflectively.
& ``empower learners to test their prompts on varying inputs, refine their templates iteratively, and engineer prompts collaboratively'' \cite{aveni_supporting_2025} \\ \hline

\makecell{\textbf{Content} \\ \textbf{Generation}}
& \Ptwelve
& Learning Material \newline Generation
& Uses AI to generate exercises, examples, explanations, prompts, practice problems, or other instructional materials for learning activities.
& ``generates programming tasks with all components necessary'' \cite{jacobs_unlimited_2025} \\ \hline

\end{tabular}
\end{table*}

Across the 90 reviewed systems, policy was rarely represented by a single isolated rule. Studies contained an average of 3.42 policy labels per system, with 85 studies (94.4\%) containing two or more policy categories. This suggests that policy was commonly expressed as a set of assistance boundaries rather than a one-dimensional allow-or-restrict decision.

Policy categories were distributed across response constraints, cognitive scaffolding, and learner adaptation~(See Table~\ref{tab:policy-definitions}). The most common individual policies were Learning-Based Personalization~(\Pnine, 51 studies, 56.7\%), Scaffolding-Oriented Support~(\Pthree, 48 studies, 53.3\%), Direct-Solution Prohibition~(\Ptwo, 38 studies, 42.2\%), Scope Restriction~(\Pone, 38 studies, 42.2\%), and Pedagogical Alignment~(\Pfive, 34 studies, 37.8\%). Overall, policy design was concentrated around response constraints, scaffolding, and personalization, while collaboration, motivation, AI literacy, and learning-material generation appeared less often.

\subsection{Enforcement}
\begin{table*}[!t]
\centering
\footnotesize
\renewcommand{\arraystretch}{1.5}
\setlength{\tabcolsep}{6pt}

\caption{Enforcement categories}
\label{tab:enforcement-definitions}

\begin{tabular}{>{\centering\arraybackslash}m{0.13\textwidth} >{\centering\arraybackslash}m{0.033\textwidth} m{0.13\textwidth} m{0.34\textwidth} m{0.24\textwidth}}
\hline
\textbf{Theme} & \textbf{ID} & \textbf{Enforcement} & \textbf{Definition} & \textbf{Representative Quote} \\ \hline

\multirow{2}{*}{\makecell{\textbf{Prompt \&} \\ \textbf{Model Control}}}
& \Eone
& Prompt-Based \newline Constraint
& Enforces boundaries through system prompts, prompt templates, role instructions, personas, or other instruction-layer constraints.
& ``The prompts include `help explain the code,' `annotate the code line by line,' `help fix the code,' and 'AI Chat.''' \cite{ez-zaouia_codigen_nodate} \\ \cline{2-5}

& \Etwo
& Backend Model or \newline Architecture Control
& Enforces boundaries through fine-tuned models, specialized classifiers, constrained decoding, backend modules, or generation pipelines beyond prompt instructions alone.
& ``The system uses reinforcement learning to continuously optimize recommendation strategies.'' \cite{gong_development_2025} \\ \hline

\multirow{2}{*}{\makecell{\textbf{Knowledge \&} \\ \textbf{Content Grounding}}}
& \Ethree
& Retrieval \newline Grounding
& Constrains AI responses by grounding them in approved materials such as course content, documentation, FAQs, syllabi, or reference resources.
& ``We retrieve k documents on each Standalone Question query.'' \cite{frazier_customizing_2024} \\ \cline{2-5}

& \Efour
& Correctness and \newline Verification Check
& Enforces boundaries by evaluating the correctness, validity, quality, or policy compliance of AI outputs or learner responses through tests, grading, filtering, validation, or regeneration.
& ``incorporating task details, grading criteria, learner solutions, and custom instructions into the prompt'' \cite{solch_direct_2025} \\ \hline

\multirow{3}{*}{\makecell{\textbf{Workflow \&} \\ \textbf{Access Control}}}
& \Efive
& Phase-Based Support \newline Workflow
& Enforces boundaries by assigning AI support to specific phases of a larger learning activity, such as preparation, independent work, feedback review, revision, or reflection.
& ``we divided the entire PBL module into three checkpoints'' \cite{kusam_generative-ai_2024} \\ \cline{2-5}

& \Esix
& Rule-Based \newline Gatekeeping
& Enforces boundaries through explicit rules, thresholds, permissions, or access conditions that allow, block, delay, or modify AI support.
& ``could think and solve the movement for themself first, and then another prompt was designed to give out the solution to the earlier'' \cite{abolnejadian_leveraging_2024} \\ \cline{2-5}

& \Eseven
& Routing-Based \newline Control
& Enforces boundaries by directing requests to different prompts, tools, models, workflows, or response strategies based on task type, query type, learner state, or error category.
& ``distinguishing between two modes: submissions that pass all test cases and those that fail some or all test cases'' \cite{nieto-cardenas_owlgorithm_2026} \\ \hline

\multirow{2}{*}{\makecell{\textbf{Interface \&} \\ \textbf{Interaction Design}}}
& \Eeight
& Interface-Level \newline Constraint
& Enforces boundaries through concrete user-interface or interaction features, such as buttons, menus, forms, locked controls, structured input fields, or predefined action options.
& ``combined with the mandatory visualization of instructional material'' \cite{aimicheva_digital_2025} \\ \cline{2-5}

& \Enine
& State Tracking and \newline Progress Control
& Uses tracked learner progress, task state, prior attempts, or interaction history to shape, limit, unlock, or display available AI support.
& ``These levels are cumulative, with each level building on the previous one as a foundation for continuous, rational learning.'' \cite{atta_smart_2025} \\ \hline

\multirow{2}{*}{\makecell{\textbf{Pedagogical} \\ \textbf{Oversight}}}
& \Eten
& Human \newline Oversight
& Enforces boundaries through instructor, TA, peer, or reviewer monitoring, approval, intervention, escalation, or feedback.
& ``Instructors and TAs are identified automatically by LTI'' \cite{liffiton_codehelp_2023} \\ \cline{2-5}

& \Eeleven
& Learner Response \newline Quality Control
& Enforces boundaries by detecting, flagging, restricting, or reviewing direct-answer submission, plagiarism, or other disallowed forms of completion.
& ``If prohibited code is detected, the system automatically sends a message instructing the learner to revise their submission'' \cite{zampirolli_intelligent_2026} \\ \hline

\end{tabular}
\end{table*}
Enforcement categories were distributed across system-level control, backend, frontend, and intervention structures~(See Table~\ref{tab:enforcement-definitions}).
Enforcement was also commonly represented by multiple codes. Studies contained an average of 2.98 enforcement labels per system, and 82 studies (91.1\%) used two or more enforcement mechanisms.

Prompt-Based Constraint~(\Eone) appeared in 82 studies (91.1\%), making it the dominant enforcement mechanism. Other mechanisms appeared less frequently, including Rule-Based Gatekeeping~(\Esix, 28 studies, 31.1\%), Backend Model or Architecture Control~(\Etwo, 28 studies, 31.1\%), Routing-Based Control~(\Eseven, 25 studies, 27.8\%), Correctness and Verification Check~(\Efour, 21 studies, 23.3\%), Interface-Level Constraint~(\Eeight, 20 studies, 22.2\%), and Retrieval Grounding~(\Ethree, 17 studies, 18.9\%). Mechanisms involving stronger runtime monitoring or pedagogical intervention were less common: Human Oversight~(\Eten) appeared in 15 studies (16.7\%), Phase-Based Support Workflow~(\Efive) and State Tracking and Progress Control~(\Enine) each appeared in 14 studies (15.6\%), and Learner Response Quality Control~(\Eeleven) appeared in 4 studies (4.4\%). Thus, although systems often used multiple enforcement mechanisms, enforcement remained strongly centered on prompt-based control.

\subsection{Authority}
\begin{table}[t]
\caption{Authority Categories}
\label{tab:authority_categories}
\centering
\footnotesize
\begin{tabular}{p{0.02\columnwidth} p{0.18\columnwidth} p{0.48\columnwidth} p{0.12\columnwidth}}
\toprule
\textbf{ID} & \textbf{Authority} & \textbf{Definition} & \textbf{Count (\%)} \\
\midrule
A1 
& System only
& The system enforces assistance boundaries, and there is no explicit evidence that learners or instructors can directly modify those boundaries through runtime system features during use. 
& 69 (76.7\%) \\
A2 
& System \newline+ learner 
& Learners can directly adjust, choose, or override assistance boundaries through explicit runtime system features during use. 
& 9 (10.0\%) \\
A3 
& System \newline+ instructor 
& Instructors can directly adjust, choose, or override assistance boundaries through explicit runtime system features during use. 
& 11 (12.2\%) \\
A4 
& System \newline + learner \newline + instructor 
& Both learners and instructors can directly adjust, choose, or override assistance boundaries through explicit runtime system features during use. 
& 1 (1.1\%) \\
\bottomrule
\end{tabular}
\end{table}
Authority is combination of system, learner and instructor~(See Table~\ref{tab:authority_categories}).
Authority was the most concentrated PEA dimension. Specifically, 69 studies (76.7\%) were coded as System only. Systems with instructor-side runtime authority appeared in 11 studies (12.2\%), and systems with learner-side runtime authority appeared in 9 studies (10.0\%). Only 1 study (1.1\%) provided evidence of both learner and instructor authority over assistance boundaries during use. Thus, while policy and enforcement often involved multiple categories, authority was usually centralized in the system.

\subsection{Policy-Enforcement Pairings}
\begin{figure}
    \centering
    \includegraphics[width=\linewidth]{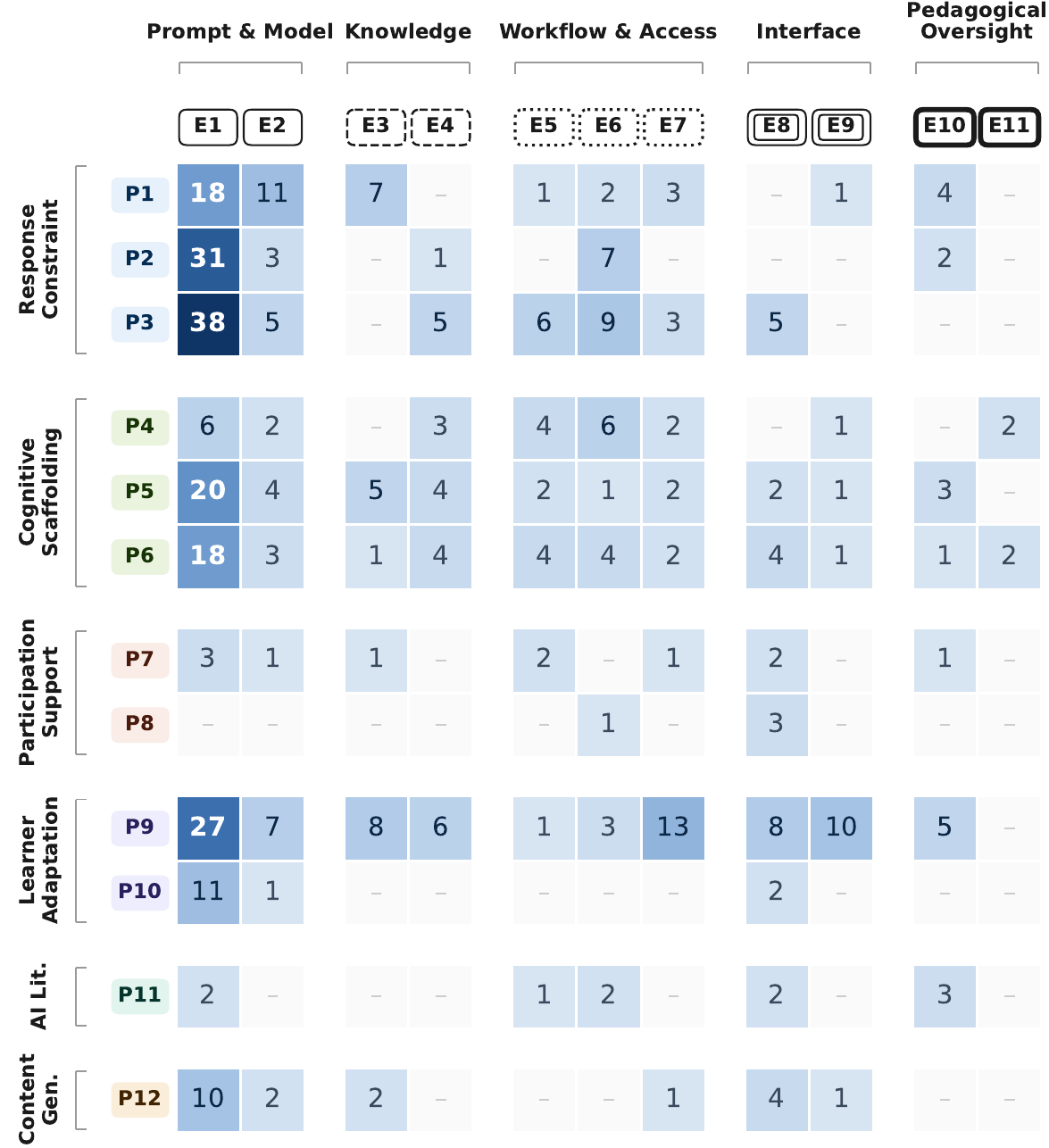}
    \caption{Policy-Enforcement Connection Heatmap}
    \label{fig:heatmap}
\end{figure}
Examining policy and enforcement together revealed how assistance boundaries were connected to implementation mechanisms (Fig.~\ref{fig:heatmap}).

\subsubsection{Prompt-based constraints dominate pairings}
Across the 12 policy categories and 11 enforcement categories, 82 of 132 possible policy--enforcement combinations were observed, leaving 50 combinations (37.9\%) absent from the corpus. The most frequent combinations all involved Prompt-Based Constraint~(\Eone): Scaffolding-Oriented Support~(\Pthree, 38 instances), Direct-Solution Prohibition~(\Ptwo, 31 instances), Learning-Based Personalization~(\Pnine, 27 instances), Pedagogical Alignment~(\Pfive, 20 instances), Demonstrating Understanding~(\Psix, 18 instances), and Scope Restriction~(\Pone, 18 instances). Prompt-Based Constraint~(\Eone) accounted for 184 of 423 observed policy--enforcement pair instances (43.5\%).

\subsubsection{Policy categories differ in enforcement breadth}

Policy categories varied in the extent to which they were put into practice through different enforcement mechanisms. Direct-Solution Prohibition~(\Ptwo) was connected to 5 enforcement mechanisms, mainly Prompt-Based Constraint~(\Eone) and Rule-Based Gatekeeping~(\Esix). Scaffolding-Oriented Support~(\Pthree) was connected to 7 mechanisms, including prompts, rules, workflows, interface constraints, verification checks, backend controls, and routing. By contrast, Learning-Based Personalization~(\Pnine) was connected to 10 mechanisms, including Prompt-Based Constraint~(\Eone), Routing-Based Control~(\Eseven), State Tracking and Progress Control~(\Enine), Retrieval Grounding~(\Ethree), Interface-Level Constraint~(\Eeight), Backend Model or Architecture Control~(\Etwo), and Correctness and Verification Check~(\Efour). Demonstrating Understanding~(\Psix) showed the broadest enforcement breadth, appearing with all 11 enforcement mechanisms. These patterns show that some policy goals were implemented through relatively narrow mechanism sets, whereas others were operationalized through a wider range of technical and interactional controls.

\subsubsection{Policy and enforcement counts were not strongly associated}

We examined whether systems with more policy categories also tended to use more enforcement mechanisms. The number of policy labels per study was not significantly associated with the number of enforcement labels by Spearman rank correlation~($\rho = .114$, $p = .283$). Thus, although many systems used multiple policy and enforcement labels, the CSV data do not support a strong tendency for systems with more policy categories to also use more enforcement mechanisms.

\subsection{Theme-Wide Policy-Enforcement Comparison}

To read the category-level policy--enforcement heatmap (Fig.~\ref{fig:heatmap}) at a broader level, we grouped the 12 policy categories and 11 enforcement categories into their larger themes. This comparison helps distinguish sparse combinations that may matter educationally from category-level absences that may simply reflect highly specific pairings.

\subsubsection{Theme-level enforcement remains centered on prompts}

Across 423 policy--enforcement pair instances, Prompt and Model Control~(\Eone--\Etwo) made up 223 instances (52.7\%), followed by Workflow and Access Control~(\Efive--\Eseven; 83 instances, 19.6\%), Knowledge and Content Grounding~(\Ethree--\Efour; 47 instances, 11.1\%), Interface and Interaction Design~(\Eeight--\Enine; 47 instances, 11.1\%), and Pedagogical Oversight~(\Eten--\Eeleven; 23 instances, 5.4\%). Thus, enforcement was still centered on prompt- and model-level control even after grouping categories into themes.

\subsubsection{Policy themes use different enforcement methods}

The pattern differed by policy theme. Response Constraint~(\Pone--\Pthree) was the largest policy theme, with 162 pair instances. Within this theme, 106 instances (65.4\%) were implemented through Prompt and Model Control~(\Eone--\Etwo), meaning that boundaries around scope, directness, and scaffolding were most often handled through prompts, role instructions, or model-level constraints.

Cognitive Scaffolding~(\Pfour--\Psix) used a wider mix of enforcement methods. Of its 114 pair instances, 53 involved Prompt and Model Control~(\Eone--\Etwo), 27 involved Workflow and Access Control~(\Efive--\Eseven), and 17 involved Knowledge and Content Grounding~(\Ethree--\Efour). This suggests that policies related to independent problem solving, pedagogical alignment, and demonstrating understanding often needed more than prompt-level instructions alone.

Learner Adaptation~(\Pnine--\Pten) fell between these two patterns. It was often implemented through Prompt and Model Control~(\Eone--\Etwo; 46 of 102 instances), but it also appeared with Interface and Interaction Design~(\Eeight--\Enine; 20 instances), Workflow and Access Control~(\Efive--\Eseven; 17 instances), and Knowledge and Content Grounding~(\Ethree--\Efour; 14 instances). This suggests that personalization was handled not only through prompts, but also through routing, state tracking, progress control, interface options, and retrieval-based grounding.

\subsubsection{Policy and enforcement themes are related}

A chi-square analysis of the theme-level matrix showed that policy themes and enforcement themes were not distributed independently, $\chi^2(20)=60.77$, $p<.001$, Cramer's $V=.190$. Because the data are multi-label and pair instances from the same paper are not fully independent, we interpret this result as a descriptive pattern rather than a causal claim. Overall, the result suggests that different policy themes tended to be connected with different enforcement methods.

\subsubsection{Underexplored areas center on participation, AI literacy, and content generation}

The theme-level comparison also shows where underexplored areas are concentrated. Most theme-level combinations appeared at least once, but Participation Support~(\Pseven--\Peight), AI Literacy~(\Peleven), and Content Generation~(\Ptwelve) appeared much less often than the other policy themes. Participation Support~(\Pseven--\Peight) accounted for only 15 pair instances, AI Literacy~(\Peleven) for 10, and Content Generation~(\Ptwelve) for 20.

Two theme-level combinations did not appear at all: AI Literacy~(\Peleven) with Knowledge and Content Grounding~(\Ethree--\Efour), and Content Generation~(\Ptwelve) with Pedagogical Oversight~(\Eten--\Eeleven). These gaps suggest that the reviewed literature reports many ways to limit, guide, and personalize individual AI assistance, but fewer
approaches to governing collaborative learning, responsible AI use, and AI-generated learning materials.

\subsection{Diversity Across PEA Dimensions}

We used the Herfindahl--Hirschman Index (HHI), calculated as the sum of squared category shares, to summarize concentration across the three PEA dimensions. Policy showed the lowest concentration (HHI = .114), followed by Enforcement (HHI = .149), while Authority was substantially more concentrated (HHI = .613). This contrast indicates that policy and enforcement were comparatively dispersed across categories, whereas authority was concentrated in system-only control.

The connection between PEA dimensions was uneven. The corpus contained a broad set of policy categories, a narrower and prompt-centered set of enforcement practices, and a highly system-centered authority distribution. In this sense, reviewed systems often showed diverse governance at the policy and enforcement levels, but centralized governance at the authority level.

\section{Discussion}
\subsection{Design Tensions in Assistance Governance}

Our findings show that assistance governance involves recurring design tradeoffs. LLM-based programming support systems need to provide useful help, limit over-reliance on generated answers, and remain aligned with course goals \cite{prather_robots_2023}. These goals are not always satisfied by a single design choice, which helps explain why similar policy goals were enforced through different mechanisms across systems. The observed policy--enforcement pairings should be interpreted as descriptive co-occurrence patterns, not as evidence that particular configurations improve learning outcomes or are preferable to less common alternatives. Below we explore policy-enforcement pairings, discuss their potential affordances and drawbacks, and highlight tensions which exist in realizing competing outcomes.

\subsubsection{Flexibility vs. Stability}

Prompt-Based Constraint~(\Eone{}) gives designers a flexible way to revise assistance boundaries without changing the interface or system architecture. However, this flexibility can make boundaries less stable and less visible, because enforcement remains dependent on instruction-layer compliance. By contrast, Rule-Based Gatekeeping~(\Esix{}), Phase-Based Support Workflow~(\Efive{}), Correctness and Verification Check~(\Efour{}), Interface-Level Constraint~(\Eeight{}), and Human Oversight~(\Eten{}) can make boundaries more explicit and enforceable, but may also reduce open-ended help seeking by constraining when, how, or through whom learners receive support.

\subsubsection{Directness vs. Learner Effort}

Less direct assistance can preserve productive struggle~\cite{kapur2012designing}, but it may also make help slower or less useful when learners are blocked. Direct-Solution Prohibition~(\Ptwo{}) was commonly enforced through Prompt-Based Constraint~(\Eone{}) or Rule-Based Gatekeeping~(\Esix{}), suggesting that many systems treated directness as a response-level constraint. In contrast, Learning-Based Personalization~(\Pnine{}) was connected to a wider range of mechanisms, including routing, learner-state tracking, backend control, retrieval grounding, interface constraints, and verification. This contrast suggests that some governance goals can be expressed as restrictions on generated responses, whereas others require tracking learner progress, task state, or performance over time.

These tensions shift the design question from whether LLM assistance should be restricted to how restrictions are operationalized. A system that discourages direct solutions through a prompt creates a different learning condition from one that does so through a staged workflow, interface constraint, correctness check, or instructor review. Thus, assistance governance should be understood not as a single restriction decision, but as a configuration of Policy, Enforcement, and Authority.

\subsection{Underexplored Regions of the PEA Design Space}

The reviewed corpus reveals uneven development within the PEA design space. These patterns identify configurations that received limited attention in the reviewed literature, and may therefore benefit from additional investigations to understand feasibility, affordances, and drawbacks.

\subsubsection{Centralized Authority}

The clearest gap concerns Authority. While Policy and Enforcement appeared in varied combinations, control over assistance boundaries was usually centralized in system logic. Most systems did not give learners or instructors explicit runtime authority to configure, adjust, or override those boundaries, suggesting that governance was commonly treated as a design-time decision rather than something adapted during learning activity.

This authority gap matters because LLM-based programming tools must account for both instructor and learner needs \cite{wu_learner_2025, hou_bespoke_2026}. Instructors may need to align support with assignment goals, assessment rules, and classroom norms, while learners may need assistance appropriate to their current understanding. However, system-centered authority should not automatically be treated as a weakness: fixed boundaries may be appropriate when consistency, assessment integrity, or simplicity is especially important.

\subsubsection{Sparse and Absent Configurations}

The reviewed literature also concentrated more heavily on governing individual learner--AI interaction than on Participation Support~(\Pseven--\Peight), AI Literacy~(\Peleven{}), and Content Generation~(\Ptwelve{}). In particular, the corpus did not report AI Literacy combined with Knowledge and Content Grounding~(\Ethree--\Efour), or Content Generation combined with Pedagogical Oversight~(\Eten--\Eeleven).

Several factors may explain these absences. Such configurations may require technically complex integrations, configurable interfaces, or coordination among learners, instructors, and institutional policies. They may also impose greater deployment and evaluation burdens, requiring longer studies, multi-stakeholder data, and measures beyond immediate task performance. Finally, runtime governance features may have received limited research attention or may not have been described in sufficient detail to be coded. The observed gaps therefore reflect both practical design challenges and the boundaries of the reviewed literature.

Together, these patterns suggest that future work should examine how governance can support shared control, learner agency, instructor alignment, collaboration, responsible AI use, and accountable content generation, while also considering the technical and institutional costs of these configurations.

\subsection{PEA as a Design Vocabulary for Future Configurations}

PEA can serve as a design vocabulary for future LLM-based programming support systems. Instead of describing a system only as a tutor, chatbot, hint generator, or feedback tool, researchers can specify the Policy goal it encodes, the Enforcement mechanism through which that goal is enacted, and the Authority arrangement that determines who can adjust it during use.

\subsubsection{Design Walkthrough}

Consider an LLM-based programming tutor intended to support introductory learners without immediately providing complete solutions. A designer could first define the \textit{Policy} as Direct-Solution Prohibition~(\Ptwo{}) combined with Scaffolding-Oriented Support~(\Pthree{}), requiring questions, explanations, or incremental hints before directly usable code. The designer could then select \textit{Enforcement} mechanisms: Prompt-Based Constraint~(\Eone{}) could restrict complete solutions, Phase-Based Support Workflow~(\Efive{}) could increase hint specificity over time, and Rule-Based Gatekeeping~(\Esix{}) could delay access to examples until the learner has made an attempt. Finally, the designer could determine \textit{Authority}: the boundaries could remain fixed by the system, be configured by an instructor, or allow learners to choose among permitted forms of support within instructor-defined limits.

This walkthrough illustrates that the same pedagogical goal can produce different learning conditions depending on how it is enforced and who can adjust it during use.

\subsubsection{Future Design Directions}

The underexplored regions suggest several possible design directions. Collaborative help routing could answer simple questions with AI hints, route conceptual questions to peers, and escalate repeated difficulties to an instructor or TA. Grounded AI literacy could ask learners to compare generated explanations with course materials, test generated code, or justify why a proposed correction is valid. Governed content generation could allow instructors to configure learning objectives, difficulty, prerequisite concepts, and review requirements before AI-generated materials are released.

These examples show how PEA can help move from analysis to design. It can also guide evaluation, because different configurations require different evidence, including learning outcomes, help-seeking quality, peer interaction, AI literacy, instructional alignment, and teacher workload. In this way, PEA provides a vocabulary for designing and evaluating LLM-based programming support according to its intended pedagogical role.

\section{Limitations}
This study has limitations. First, the review is bounded by the published literature included in selected databases and within a defined time window, so relevant systems described in preprints, workshops, industry tools, or local course deployments may not be represented. Second, our inclusion criteria shape the observed design space. Our focus on peer-reviewed, learner-facing systems with meaningful adaptations and user studies improves comparability, but excludes industry tools, gray literature, preprints, local deployments, general-purpose LLM use, and unevaluated systems. Therefore, sparse PEA configurations should be interpreted as underexplored in this corpus, not as infeasible or absent from practice. Third, full-text exclusion reasons were not logged as mutually exclusive categories, so we report only the aggregate exclusion count. This limits the reproducibility of the screening audit trail and should be considered when interpreting the corpus construction.

\section{Conclusion}
As LLM-based programming support becomes more common in CS education, the central design question is no longer only what these systems can generate, but how their assistance is governed. This review addressed that question by introducing PEA as an analytic lens for describing assistance governance. Our findings suggest that systems in the reviewed corpus often encode similar educational intentions; however, these intentions
are operationalized through different enforcement mechanisms. The review also shows an uneven design space: while policy and enforcement appear in varied combinations, authority is usually centralized in system logic, leaving learner and instructor control comparatively underdeveloped.

PEA contributes a vocabulary for making these choices explicit. For researchers, it supports comparison beyond model capability, task coverage, or learning outcome alone. For designers and educators, it offers a way to reason more deliberately about what forms of help should be available, how assistance boundaries should be implemented, and who should be able to adjust them during learning activity. Future work can build on this foundation by testing specific PEA configurations, examining their effects in longer-term deployments, and exploring governance arrangements that better balance learning support, instructional accountability, and learner agency.

\section*{Acknowledgement}
The authors used ChatGPT, Gemini, and Claude to assist with language editing, clarity improvement, and manuscript organization. The authors reviewed, revised, and approved all AI-assisted content and take full responsibility for the final manuscript.


\clearpage
\appendix
\subsection{Keyword-Based Filtering Criteria for Secondary Screening}
\label{app:filtering-criteria}

To support secondary screening, we used a keyword-based filtering framework as a human screening guide. The keyword groups were not used as an automatic exclusion rule. Instead, reviewers used them to judge whether a record was likely to match the review scope during title and abstract screening. Records were prioritized when they included terms related to (1) large language models, (2) computer science education, (3) educational systems or tools, and (4) tool creation, modification, or design.

\subsubsection{Topic: Large Language Models}
\begin{itemize}
\item large language model
\item LLM
\item generative AI
\item GenAI
\item ChatGPT
\item GPT
\item Gemini
\item Claude
\item DeepSeek
\end{itemize}

\subsubsection{Domain: Computer Science Education}
\begin{itemize}
\item computer science education
\item computing education
\item CS education
\item CSEd
\item CER
\item computing students
\item computing instructors
\item CS students
\item CS instructors
\item computer engineering education
\item programming education
\item introductory programming
\item introductory programming course
\item novice programmers
\item novice
\item programming class
\item programming course
\item classroom
\end{itemize}

\subsubsection{Focus: Systems and Tools}
\begin{itemize}
\item chatbot
\item digital teaching assistant
\item tool
\item tools
\item system
\item systems
\item educational system
\item educational systems
\item tutorial
\item tutorials
\item assistant
\item assistants
\item platform
\item platforms
\item tutor
\item tutors
\end{itemize}

\subsubsection{Focus: Tool Creation, Modification, or Design}
\begin{itemize}
\item design
\item designed
\item create
\item created
\item introduce
\item introduced
\item develop
\item developed
\item prototype
\item prototyped
\item modify
\item modified
\end{itemize}

\subsection{Database Search Queries}
\label{app:search-query}

We report the database-specific search strings used in ACM Digital Library, IEEE Xplore, and Scopus. We set January 2023 as the starting point because ChatGPT was launched on November 30, 2022, shortly before the rapid growth of LLM-based educational tool research. Searches were conducted on March 10, 2026. Searches were applied to title and abstract fields where supported and were limited to English-language, peer-reviewed journal or conference/proceedings records in CS, computing, or engineering-related areas where such filters were available.

Because each database provides different search syntax, metadata fields, and filtering behavior, the search strings were adapted by database. In ACM Digital Library, we used a broader title/abstract query combining LLM-related terms, CS/programming education terms, and learner-facing system/tool terms. We did not include the tool creation, modification, or design terms directly in the ACM query because doing so reduced search sensitivity and removed known relevant seed papers. Instead, those terms were used during human secondary screening as part of the keyword-based screening guide. In IEEE Xplore and Scopus, the tool creation, modification, and design terms were included directly in the database query because the resulting searches preserved the relevant seed papers while improving precision.

\subsubsection{ACM Digital Library}

The ACM Digital Library search returned 1,560 records.

\begin{lstlisting}[style=query]
(Title:("large language model" OR "large language models" OR "llm" OR "llms" OR "generative ai" OR "genai" OR "chatgpt" OR "gpt" OR "gemini" OR "claude" OR "deepseek") OR Abstract:("large language model" OR "large language models" OR "llm" OR "llms" OR "generative ai" OR "genai" OR "chatgpt" OR "gpt" OR "gemini" OR "claude" OR "deepseek")) AND (Title:("computer science education" OR "computing education" OR "cs education" OR "csed" OR "cer" OR "computing students" OR "computing instructors" OR "cs students" OR "cs instructors" OR "computer engineering education" OR "programming education" OR "introductory programming" OR "novice programmer" OR "novice programmers" OR "novice" OR "novices" OR "programming class" OR "programming classes" OR "students" OR "instructors" OR "course" OR "courses" OR "classroom" OR "classrooms") OR Abstract:("computer science education" OR "computing education" OR "cs education" OR "csed" OR "cer" OR "computing students" OR "computing instructors" OR "cs students" OR "cs instructors" OR "computer engineering education" OR "programming education" OR "introductory programming" OR "novice programmer" OR "novice programmers" OR "novice" OR "novices" OR "programming class" OR "programming classes" OR "students" OR "instructors" OR "course" OR "courses" OR "classroom" OR "classrooms")) AND (Title:("chatbot" OR "chatbots" OR "digital teaching assistant" OR "digital teaching assistants" OR "tool" OR "tools" OR "system" OR "systems" OR "educational system" OR "educational systems" OR "tutorial" OR "tutorials" OR "assistant" OR "assistants" OR "platform" OR "platforms" OR "tutor" OR "tutors" OR "tutoring") OR Abstract:("chatbot" OR "chatbots" OR "digital teaching assistant" OR "digital teaching assistants" OR "tool" OR "tools" OR "system" OR "systems" OR "educational system" OR "educational systems" OR "tutorial" OR "tutorials" OR "assistant" OR "assistants" OR "platform" OR "platforms" OR "tutor" OR "tutors" OR "tutoring"))
\end{lstlisting}

\subsubsection{IEEE Xplore}

The IEEE Xplore search returned 1,143 records.

\begin{lstlisting}[style=query]
((("Abstract":"large language model" OR "Document Title":"large language model" OR
"Abstract":"large language models" OR "Document Title":"large language models" OR
"Abstract":"llm" OR "Document Title":"llm" OR "Abstract":"llms" OR "Document Title":"llms" OR
"Abstract":"generative ai" OR "Document Title":"generative ai" OR "Abstract":"genai" OR "Document Title":"genai" OR
"Abstract":"chatgpt" OR "Document Title":"chatgpt" OR "Abstract":"gpt" OR "Document Title":"gpt" OR
"Abstract":"gemini" OR "Document Title":"gemini" OR "Abstract":"claude" OR "Document Title":"claude" OR
"Abstract":"deepseek" OR "Document Title":"deepseek") AND
("Abstract":"computer science education" OR "Document Title":"computer science education" OR
"Abstract":"computing education" OR "Document Title":"computing education" OR
"Abstract":"cs education" OR "Document Title":"cs education" OR "Abstract":"csed" OR "Document Title":"csed" OR
"Abstract":"cer" OR "Document Title":"cer" OR "Abstract":"computing students" OR "Document Title":"computing students" OR
"Abstract":"computing instructors" OR "Document Title":"computing instructors" OR
"Abstract":"cs students" OR "Document Title":"cs students" OR "Abstract":"cs instructors" OR "Document Title":"cs instructors" OR
"Abstract":"computer engineering education" OR "Document Title":"computer engineering education" OR
"Abstract":"programming education" OR "Document Title":"programming education" OR
"Abstract":"introductory programming" OR "Document Title":"introductory programming" OR
"Abstract":"novice programmer" OR "Document Title":"novice programmer" OR
"Abstract":"novice programmers" OR "Document Title":"novice programmers" OR
"Abstract":"novice" OR "Document Title":"novice" OR "Abstract":"novices" OR "Document Title":"novices" OR
"Abstract":"programming class" OR "Document Title":"programming class" OR
"Abstract":"programming classes" OR "Document Title":"programming classes" OR
"Abstract":"students" OR "Document Title":"students" OR "Abstract":"instructors" OR "Document Title":"instructors" OR
"Abstract":"course" OR "Document Title":"course" OR "Abstract":"courses" OR "Document Title":"courses" OR
"Abstract":"classroom" OR "Document Title":"classroom" OR "Abstract":"classrooms" OR "Document Title":"classrooms") AND
("Abstract":"chatbot" OR "Document Title":"chatbot" OR "Abstract":"chatbots" OR "Document Title":"chatbots" OR
"Abstract":"digital teaching assistant" OR "Document Title":"digital teaching assistant" OR
"Abstract":"digital teaching assistants" OR "Document Title":"digital teaching assistants" OR
"Abstract":"tool" OR "Document Title":"tool" OR "Abstract":"tools" OR "Document Title":"tools" OR
"Abstract":"system" OR "Document Title":"system" OR "Abstract":"systems" OR "Document Title":"systems" OR
"Abstract":"educational system" OR "Document Title":"educational system" OR
"Abstract":"educational systems" OR "Document Title":"educational systems" OR
"Abstract":"tutorial" OR "Document Title":"tutorial" OR "Abstract":"tutorials" OR "Document Title":"tutorials" OR
"Abstract":"assistant" OR "Document Title":"assistant" OR "Abstract":"assistants" OR "Document Title":"assistants" OR
"Abstract":"platform" OR "Document Title":"platform" OR "Abstract":"platforms" OR "Document Title":"platforms" OR
"Abstract":"tutor" OR "Document Title":"tutor" OR "Abstract":"tutors" OR "Document Title":"tutors" OR
"Abstract":"tutoring" OR "Document Title":"tutoring") AND
("Abstract":"design" OR "Document Title":"design" OR "Abstract":"designs" OR "Document Title":"designs" OR
"Abstract":"designed" OR "Document Title":"designed" OR "Abstract":"designing" OR "Document Title":"designing" OR
"Abstract":"create" OR "Document Title":"create" OR "Abstract":"creates" OR "Document Title":"creates" OR
"Abstract":"created" OR "Document Title":"created" OR "Abstract":"creating" OR "Document Title":"creating" OR
"Abstract":"introduce" OR "Document Title":"introduce" OR "Abstract":"introduces" OR "Document Title":"introduces" OR
"Abstract":"introduced" OR "Document Title":"introduced" OR "Abstract":"introducing" OR "Document Title":"introducing" OR
"Abstract":"develop" OR "Document Title":"develop" OR "Abstract":"develops" OR "Document Title":"develops" OR
"Abstract":"developed" OR "Document Title":"developed" OR "Abstract":"developing" OR "Document Title":"developing" OR
"Abstract":"prototype" OR "Document Title":"prototype" OR "Abstract":"prototypes" OR "Document Title":"prototypes" OR
"Abstract":"prototyped" OR "Document Title":"prototyped" OR "Abstract":"prototyping" OR "Document Title":"prototyping" OR
"Abstract":"modify" OR "Document Title":"modify" OR "Abstract":"modifies" OR "Document Title":"modifies" OR
"Abstract":"modified" OR "Document Title":"modified" OR "Abstract":"modifying" OR "Document Title":"modifying")))
\end{lstlisting}

\subsubsection{Scopus}

The Scopus search returned 3,125 records.

\begin{lstlisting}[style=query]
( TITLE-ABS ( "large language model" OR "large language models" OR "llm" OR "llms" OR "generative ai" OR "genai" OR "chatgpt" OR "gpt" OR "gemini" OR "claude" OR "deepseek" ) AND TITLE-ABS ( "computer science education" OR "computing education" OR "cs education" OR "csed" OR "cer" OR "computing students" OR "computing instructors" OR "cs students" OR "cs instructors" OR "computer engineering education" OR "programming education" OR "introductory programming" OR "novice programmer" OR "novice programmers" OR "novice" OR "novices" OR "programming class" OR "programming classes" OR "students" OR "instructors" OR "course" OR "courses" OR "classroom" OR "classrooms" ) AND TITLE-ABS ( "chatbot" OR "chatbots" OR "digital teaching assistant" OR "digital teaching assistants" OR "tool" OR "tools" OR "system" OR "systems" OR "educational system" OR "educational systems" OR "tutorial" OR "tutorials" OR "assistant" OR "assistants" OR "platform" OR "platforms" OR "tutor" OR "tutors" OR "tutoring" ) AND TITLE-ABS ( "design" OR "designs" OR "designed" OR "designing" OR "create" OR "creates" OR "created" OR "creating" OR "introduce" OR "introduces" OR "introduced" OR "introducing" OR "develop" OR "develops" OR "developed" OR "developing" OR "prototype" OR "prototypes" OR "prototyped" OR "prototyping" OR "modify" OR "modifies" OR "modified" OR "modifying" ) AND PUBYEAR > 2022 AND PUBYEAR < 2027 ) AND ( LIMIT-TO ( SUBJAREA , "COMP" ) OR LIMIT-TO ( SUBJAREA , "ENGI" ) ) AND ( LIMIT-TO ( DOCTYPE , "cp" ) OR LIMIT-TO ( DOCTYPE , "ar" ) ) AND ( LIMIT-TO ( LANGUAGE , "English" ) ) AND ( LIMIT-TO ( SRCTYPE , "p" ) OR LIMIT-TO ( SRCTYPE , "j" ) ) AND ( LIMIT-TO ( PUBSTAGE , "final" ) )
\end{lstlisting}

\subsection{Screening, Deduplication, and Audit Trail}
\label{app:screening-audit}

All records retrieved from ACM Digital Library, IEEE Xplore, and Scopus were imported into Rayyan for screening. Duplicate records were first identified using Rayyan's automatic duplicate-detection function based primarily on DOI and title metadata. The automatically identified duplicates were then manually reviewed. Ambiguous cases, including records with similar titles or incomplete metadata, were checked manually before removal. This process removed 93 duplicate records.

The keyword groups listed above were used as a human screening guide rather than as an automatic exclusion rule. Reviewers used these terms to support title and abstract screening and to judge whether a record was likely to match the review scope. Records were assessed against the inclusion criteria described in the main paper, including whether the paper reported a learner-facing LLM-based programming support tool or prototype, was situated in CS or programming education, included a meaningful adaptation beyond direct use of a general-purpose LLM interface, included a user study, and provided sufficient detail for PEA extraction.

During full-text screening, papers were excluded when they did not report a learner-facing tool, system, or prototype; used ChatGPT or another general-purpose LLM interface without meaningful adaptation; were not situated in CS or programming education; did not include a user study; focused only on course, curriculum, or instructional planning; provided insufficient system detail for PEA extraction; or were under five pages. We report the total number of excluded full-text papers in the PRISMA flow diagram. Exclusion reasons were applied during screening, but they were not logged as mutually exclusive counts by reason; therefore, we do not report a separate exclusion-count table by reason.




\subsection{LLM-Assisted Extraction and Coding}
\label{app:llm-agent}

LLM-assisted extraction and coding support was conducted between March and April 2026. We used LLM tools only as supportive aids during preliminary extraction and coding. Descriptive extraction was supported using both a custom GPT in ChatGPT and a Claude Project. Qualitative Policy--Enforcement coding support was conducted using the custom GPT in ChatGPT. Exact backend model versions were visible in the interfaces at the time of use but were not separately logged. We used the default interface settings and did not manually configure temperature, seed, or top-p.

The LLM tools were not treated as independent coders, raters, or adjudicators. Their outputs were used only as provisional suggestions to help identify candidate evidence, possible descriptive fields, and possible Policy--Enforcement annotations. Descriptive extraction outputs were checked against the source papers and entered directly into the extraction spreadsheet. Qualitative coding outputs were recorded in internal Notion notes. The LLM outputs included candidate evidence quotes and rationales, and human researchers checked the quoted evidence against the full source papers before using it. No LLM-generated label was accepted automatically. All labels were reviewed against the finalized codebook, revised or rejected when necessary, and supplemented by human researchers when relevant evidence had been missed.

Authority coding was conducted as part of the descriptive analysis workflow. Authority categories were assigned based on whether the paper reported explicit runtime system features through which learners or instructors could adjust assistance boundaries. The four Authority categories were System only, System + learner, System + instructor, and System + learner + instructor.

\subsubsection{Descriptive Extraction Prompt Schema}

The following prompt schema was used to support descriptive extraction for each paper. The model was asked to identify candidate evidence and suggest values for predefined descriptive fields. For closed-ended fields, the model was asked to select the closest available category and justify the selection. If no category fit, any suggested new category was treated as a flag for human review rather than accepted automatically.

\begin{lstlisting}[style=query]
Pretend you are a researcher conducting a careful review of a paper.

Your task is to extract candidate evidence for the following descriptive criteria:
<insert all descriptive columns, definitions, and whether each field is closed-ended or open-ended. For closed-ended fields, include the allowed categories.>

For each criterion:

1. Extract relevant quotes or passages from the paper.
2. Based on the evidence, suggest the appropriate field value.
3. If the criterion is closed-ended, select the closest allowed category and briefly justify the selection.
4. If none of the allowed categories fit, flag the case for human review and explain why.
5. If the criterion is open-ended, propose a concise descriptive value and provide a short definition if needed.

Return the output in a structured table with the following columns:
Criterion | Candidate value | Evidence quote(s) | Rationale | Needs human review?
\end{lstlisting}

\subsubsection{Qualitative Coding Support Prompt Schema}

The following prompt schema was used during preliminary qualitative coding before the final PEA codebook was consolidated. The model outputs were later reviewed, revised, or rejected by human researchers and mapped to the finalized codebook.

\begin{lstlisting}[style=query]
Pretend you are a researcher supporting qualitative coding for a scoping review.

We are analyzing assistance governance in learner-facing LLM-based programming support systems. Identify candidate Policy and Enforcement evidence from the provided paper.

Definitions:

* Policy: what forms of assistance are educationally allowed, restricted, delayed, encouraged, or prohibited.
* Enforcement: how those assistance boundaries are implemented through prompts, interface features, workflows, rules, model architecture, retrieval, verification, oversight, or other system behavior.

Scope limitations:

* Analyze only what the system actually implements, configures, or makes available to users.
* Include prototype or experimental-system features when the paper states that they were implemented.
* Exclude research methodology, experimental design, survey items, interview protocols, participant perceptions, author interpretations, design implications, future proposals, and recommendations.
* Exclude conceptual proposals whose implementation status is unclear.
* Use only evidence from the paper. Do not infer a code unless the paper provides textual evidence.
* If the evidence is ambiguous, flag the case for human review rather than over-interpreting it.

Policy inclusion guidance:

* Include content related to learning, teaching, tutoring, feedback, scaffolding, assessment, assignment completion, student support, class management, or educational interaction.
* Include rules that influence educational behavior, such as prohibiting direct answers, providing hints only, restricting help to course-related questions, adapting difficulty, or requiring citation/grounding for learning support.
* Exclude general product rules, operational settings, or platform safety constraints unless they shape educational assistance.

Enforcement inclusion guidance:

* Include implementation mechanisms such as buttons, menus, dropdowns, branched workflows, role prompts, prompt templates, retrieval pipelines, classifiers, validators, gating logic, logging, moderation filters, memory stores, LMS integration, backend rules, and human review.
* Do not classify a simple functional description as Enforcement unless it shows how a Policy is implemented.
* When possible, identify which Enforcement mechanism operationalizes which Policy.

For each candidate Policy or Enforcement code:

1. Provide the dimension: Policy or Enforcement.
2. Provide a provisional code name.
3. Provide the exact evidence quote or passage.
4. Provide the page number if available.
5. Explain why the evidence supports the provisional code.
6. Indicate whether the evidence is explicit or implicit.
7. Flag uncertainty or possible alternative interpretations.

Return the output in a structured table with the following columns:
Dimension | Provisional code | Evidence quote(s) | Page | Rationale | Explicit/Implicit | Uncertainty or notes
\end{lstlisting}

\subsection{Paper-Level Coding Tables}
\label{app:paper-level-coding}

The following paper-level coding tables provide the coding results used in the analysis. The first table reports the Policy--Enforcement pairs assigned to each reviewed paper. The second table reports the corresponding Authority category and Authority code. These tables provide paper-level coding transparency for the reported PEA results. Quote-level evidence and rationales were maintained internally during coding but are not included in the supplement due to space constraints.
\begin{table}[!t]
\centering
\scriptsize
\renewcommand{\arraystretch}{0.82}
\setlength{\tabcolsep}{2.3pt}
\setlength{\abovecaptionskip}{2pt}
\setlength{\belowcaptionskip}{3pt}
\caption{Paper--authority mapping}
\label{tab:paper-authority-mapping}
\begin{tabular}{@{}>{\raggedright\arraybackslash}p{0.1\columnwidth}>{\raggedright\arraybackslash}p{0.34\columnwidth}c@{}}
\toprule
\textbf{Paper} & \textbf{Authority} & \textbf{Code} \\
\midrule
\cite{zamfirescu-pereira_61a_2025} & System & A1 \\
\cite{aimicheva_digital_2025} & System & A1 \\
\cite{sooriamurthi_generative_2025} & System & A1 \\
\cite{qi_knowledge-component-based_2025} & System & A1 \\
\cite{kweon_large-scale_nodate} & System & A1 \\
\cite{atta_smart_2025} & System & A1 \\
\cite{woodrow_ai_2024} & System & A1 \\
\cite{dickey_ai-lab_2024} & System & A1 \\
\cite{tan_bridging_2025} & System & A1 \\
\cite{kazemitabaar_codeaid_2024} & System, Student & A2 \\
\cite{liffiton_codehelp_2023} & System, Teacher & A3 \\
\cite{alrabah_codelens_2025} & System & A1 \\
\cite{hou_codetailor_2024} & System, Student & A2 \\
\cite{ez-zaouia_codigen_nodate} & System & A1 \\
\cite{chen_coding_2024} & System & A1 \\
\cite{renzella_compiler-integrated_2025} & System & A1 \\
\cite{frazier_customizing_2024} & System & A1 \\
\cite{ma_dbox_2025} & System & A1 \\
\cite{taylor_dcc_2024} & System & A1 \\
\cite{yang_debugging_2024} & System & A1 \\
\cite{petula_deliverc_2026} & System & A1 \\
\cite{wang_devcoach_2025} & System & A1 \\
\cite{yosobumi_development_2024} & System & A1 \\
\cite{gong_development_2025} & System & A1 \\
\cite{solch_direct_2025} & System, Teacher & A3 \\
\cite{martinez_enhancing_2025} & System & A1 \\
\cite{ramasamy_enhancing_2024} & System & A1 \\
\cite{chrysopoulos_enhancing_2025} & System, Teacher & A3 \\
\cite{kimmel_enhancing_2024} & System & A1 \\
\cite{yang_enhancing_2024} & System & A1 \\
\cite{vemula_enriching_2024} & System & A1 \\
\cite{logacheva_evaluating_2024} & System & A1 \\
\cite{noraset_evaluating_2026} & System, Student & A2 \\
\cite{jury_evaluating_2024} & System, Student & A2 \\
\cite{zhou_expeerience_2026} & System & A1 \\
\cite{hoq_explainable_nodate} & System, Teacher & A3 \\
\cite{denny_explaining_2024} & System & A1 \\
\cite{zhao_exploration_2025} & System, Teacher & A3 \\
\cite{stienstra_exploring_2025} & System & A1 \\
\cite{xiao_exploring_2024} & System, Student & A2 \\
\cite{khor_exploring_2025} & System & A1 \\
\cite{liu_fastfixer_2024} & System & A1 \\
\cite{zhu_feedbot_2025} & System & A1 \\
\cite{ouaazki_generative_2024} & System & A1 \\
\cite{kusam_generative-ai_2024} & System, Student & A2 \\
\cite{liu_improving_2025} & System, Teacher & A3 \\
\cite{meza_improving_2024} & System & A1 \\
\cite{shochcho_improving_2025} & System & A1 \\
\cite{li_integrating_2025} & System & A1 \\
\cite{zampirolli_intelligent_2026} & System & A1 \\
\cite{chandarana_investigating_2025} & System & A1 \\
\cite{santos_its_2025} & System & A1 \\
\cite{kuramitsu_kogi_2023} & System & A1 \\
\cite{takamatsu_enhancing_2025} & System, Student & A2 \\
\cite{bassner_less_2026} & System & A1 \\
\cite{abolnejadian_leveraging_2024} & System & A1 \\
\cite{yang_leveraging_2026} & System & A1 \\
\cite{youn_llm-based_2025} & System & A1 \\
\cite{rachha_llm-enhanced_2024} & System & A1 \\
\cite{pankiewicz_navigating_2024} & System & A1 \\
\cite{roest_next-step_2024} & System & A1 \\
\cite{santos_not_2024} & System & A1 \\
\cite{su_oi-assistant_2025} & System & A1 \\
\cite{hassan_teaching_2025} & System & A1 \\
\cite{birillo_one_2024} & System, Student & A2 \\
\cite{nieto-cardenas_owlgorithm_2026} & System & A1 \\
\cite{tran_pacing_2026} & System, Teacher & A3 \\
\cite{troussas_personagpt_2025} & System & A1 \\
\cite{rogers_playing_2025} & System & A1 \\
\cite{padurean_prompt_2025} & System & A1 \\
\cite{perez_prompting_2025} & System & A1 \\
\cite{zhang_prompttutor_2025} & System & A1 \\
\cite{kelly_quack_2025} & System & A1 \\
\cite{smith_redefining_2025} & System & A1 \\
\cite{liao_scaffolding_2024} & System, Teacher & A3 \\
\cite{wiktor_supporting_2026} & System & A1 \\
\cite{aveni_supporting_2025} & System, Student, Teacher & A4 \\
\cite{alario-hoyos_tailoring_2024} & System & A1 \\
\cite{jin_teach_2024} & System, Teacher & A3 \\
\cite{liu_teaching_2024} & System, Teacher & A3 \\
\cite{jacobs_thats_2025} & System & A1 \\
\cite{mueller_power_2025} & System & A1 \\
\cite{jell_towards_2023} & System, Teacher & A3 \\
\cite{wu_tracemate_2025} & System & A1 \\
\cite{kuramitsu_training_2024} & System & A1 \\
\cite{jacobs_unlimited_2025} & System & A1 \\
\cite{qu_unlocking_2024} & System & A1 \\
\cite{zhu_unrealmentor_2025} & System & A1 \\
\cite{yarleque_web-based_2025} & System & A1 \\
\cite{richards_what_2024} & System, Student & A2 \\
\bottomrule
\end{tabular}
\end{table}

%

\begin{table*}[!t]
\centering
\fontsize{6.7pt}{6.2pt}\selectfont
\renewcommand{\arraystretch}{0.56}
\setlength{\tabcolsep}{1.5pt}
\setlength{\abovecaptionskip}{2pt}
\setlength{\belowcaptionskip}{3pt}

\caption{Reviewed papers and coded Policy--Enforcement--Authority categories}
\label{tab:paper-pea-coding}

\makebox[\textwidth][c]{%
\begin{tabular}{@{}>{\raggedright\arraybackslash}p{0.69\textwidth}@{\hspace{7pt}}>{\raggedright\arraybackslash}p{0.28\textwidth}@{\hspace{6pt}}>{\raggedright\arraybackslash}p{0.04\textwidth}@{}}
\toprule
\textbf{TITLE} & \textbf{POLICY--ENFORCEMENT} & \textbf{CITE} \\
\midrule
61A Bot Report: AI Assistants in CS1 Save Students Homework Time and Reduce Demands on Staff. (Now What?)
& P2--E1, P4--E6, P3--E1, P6--E1
& \cite{zamfirescu-pereira_61a_2025} \\ 

A Digital Ecosystem Model for Developing Logical Thinking in Novice Programmers: Integrating Visualization Technologies and GenAI
& P6--E1, P3--E5, P3--E8, P9--E5
& \cite{aimicheva_digital_2025} \\ 

A Generative AI Tool to Foster and Assess Authentic Learning: A Case Study in Teaching SQL
& P2--E1, P6--E1, P6--E2, P6--E6, P6--E3
& \cite{sooriamurthi_generative_2025} \\ 

A Knowledge-Component-Based Methodology for Evaluating AI Assistants
& P3--E1, P1--E7, P3--E8
& \cite{qi_knowledge-component-based_2025} \\ 

A Large-Scale Real-World Evaluation of an LLM-Based Virtual Teaching Assistant
& P1--E1, P1--E3, P1--E10
& \cite{kweon_large-scale_nodate} \\ 

A Smart ChatGPT Mobile Application for Improving C\_\_HASHESC\_\_ Programming Skills for Students in Educational Institutions
& P4--E6, P2--E6, P4--E9, P5--E5, P9--E4, P9--E9, P5--E8
& \cite{atta_smart_2025} \\ 

AI Teaches the Art of Elegant Coding: Timely, Fair, and Helpful Style Feedback in a Global Course
& P1--E6, P1--E1, P4--E6, P3--E1, P3--E4
& \cite{woodrow_ai_2024} \\ 

AI-Lab: A Framework for Introducing Generative Artificial Intelligence Tools in Computer Programming Courses
& P3--E5, P11--E1, P4--E5, P4--E1, P6--E1, P6--E10
& \cite{dickey_ai-lab_2024} \\ 

Bridging the AI-Human Gap in Programming Education: A Case Study on Structured Prompting and Content-Constrained Generation
& P1--E3, P5--E1, P9--E1, P5--E3, P5--E7, P9--E3, P4--E1
& \cite{tan_bridging_2025} \\ 

CodeAid: Evaluating a Classroom Deployment of an LLM-based Programming Assistant that Balances Student and Educator Needs
& P2--E1, P9--E7, P10--E8, P9--E8, P1--E1, P9--E3
& \cite{kazemitabaar_codeaid_2024} \\ 

CodeHelp: Using Large Language Models with Guardrails for Scalable Support in Programming Classes
& P2--E1, P1--E1, P6--E6, P11--E10, P2--E4
& \cite{liffiton_codehelp_2023} \\ 

CodeLens: A Generative AI Framework for Automated Feedback on SQL Assignments
& P2--E1, P4--E2, P3--E1, P3--E4
& \cite{alrabah_codelens_2025} \\ 

CodeTailor: LLM-Powered Personalized Parsons Puzzles for Engaging Support While Learning Programming
& P9--E1, P12--E2, P7--E2, P12--E8, P3--E8, P9--E4
& \cite{hou_codetailor_2024} \\ 

Codigen: Enhancing Personalized Programming Education with Visual Authoring and LLMs
& P12--E8, P5--E8, P2--E1, P6--E8
& \cite{ez-zaouia_codigen_nodate} \\ 

Coding Error-Based Reflective Learning with ChatGPT: A Study on Higher-order Thinking Skills in Programming Education
& P4--E6, P6--E1, P3--E1, P3--E2, P9--E9
& \cite{chen_coding_2024} \\ 

Compiler-Integrated, Conversational AI for Debugging CS1 Programs
& P9--E1, P9--E9, P9--E10, P2--E1, P2--E6, P3--E1, P9--E8
& \cite{renzella_compiler-integrated_2025} \\ 

Customizing ChatGPT to Help Computer Science Principles Students Learn Through Conversation
& P9--E1, P5--E3, P6--E1, P12--E1
& \cite{frazier_customizing_2024} \\ 

DBox: Scaffolding Algorithmic Programming Learning through Learner-LLM Co-Decomposition
& P4--E5, P6--E8, P6--E4, P3--E6, P3--E1, P4--E7
& \cite{ma_dbox_2025} \\ 

dcc --help: Transforming the Role of the Compiler by Generating Context-Aware Error Explanations with Large Language Models
& P9--E1, P2--E1, P11--E8, P11--E6
& \cite{taylor_dcc_2024} \\ 

Debugging with an AI Tutor: Investigating Novice Help-seeking Behaviors and Perceived Learning
& P5--E1, P2--E1, P1--E3
& \cite{yang_debugging_2024} \\ 

DeliverC: Teaching Pointers through GenAI-Powered Game-Based Learning
& P9--E1, P12--E1, P9--E7, P4--E1, P9--E8
& \cite{petula_deliverc_2026} \\ 

DevCoach: Supporting Students Learning the Software Development Life Cycle with a Generative AI Powered Multi-Agent System
& P7--E1, P2--E1, P3--E1, P6--E8, P7--E5, P4--E1, P6--E5
& \cite{wang_devcoach_2025} \\ 

Development of a Pedagogical Agent Utilizing ChatGPT as a Response Mechanism and Scaffolding Method for CSS Learning
& P3--E1, P3--E5, P7--E1, P9--E6, P9--E1, P4--E5, P6--E5, P3--E6
& \cite{yosobumi_development_2024} \\ 

Development of Learning Agent for Data Structures on Smart Education Platform
& P9--E9, P9--E6, P9--E2, P5--E3, P3--E1, P9--E3, P9--E1
& \cite{gong_development_2025} \\ 

Direct Automated Feedback Delivery for Student Submissions based on LLMs
& P9--E9, P2--E1, P5--E1
& \cite{solch_direct_2025} \\ 

Enhancing Algorithmic Thinking and Emotional Resilience in Programming Education Through AI Powered Virtual Tutoring
& P3--E5, P6--E6, P2--E1, P10--E1, P7--E5, P6--E5
& \cite{martinez_enhancing_2025} \\ 

Enhancing CS Education with LAs Using AI-Empowered AIELA Program
& P1--E1, P7--E10, P7--E7, P3--E1, P9--E10, P2--E1
& \cite{ramasamy_enhancing_2024} \\ 

Enhancing High School Programming Education Through LLM-Based Hint Generation
& P3--E1, P5--E1, P5--E10, P9--E10, P1--E2, P1--E10
& \cite{chrysopoulos_enhancing_2025} \\ 

Enhancing Programming Error Messages in Real Time with Generative AI
& P4--E4, P2--E1
& \cite{kimmel_enhancing_2024} \\ 

Enhancing python learning with PyTutor: Efficacy of a ChatGPT-Based intelligent tutoring system in programming education
& P3--E6, P3--E1, P2--E6, P11--E10, P5--E1
& \cite{yang_enhancing_2024} \\ 

Enriching Python Programming Education With Generative AI: Leveraging Large Language Models for Personalized Support and Interactive Learning
& P1--E2, P5--E1
& \cite{vemula_enriching_2024} \\ 

Evaluating Contextually Personalized Programming Exercises Created with Generative AI
& P12--E8, P8--E8, P9--E8, P12--E1, P1--E1
& \cite{logacheva_evaluating_2024} \\ 

Evaluating lab assistant chatbot on student learning and behaviors in a programming short course
& P2--E1, P9--E1, P3--E7
& \cite{noraset_evaluating_2026} \\ 

Evaluating LLM-generated Worked Examples in an Introductory Programming Course
& P5--E1, P12--E7, P12--E1, P12--E8, P9--E8
& \cite{jury_evaluating_2024} \\ 

ExPeerience: Towards AI-Assisted Learnersourcing to Bridge Conceptual Understanding and Problem Solving in Database Programming Education
& P3--E5, P12--E1, P9--E1, P3--E1, P6--E4, P9--E7, P6--E11
& \cite{zhou_expeerience_2026} \\ 

Explainable AI in the Loop: An Instructor-Transformer Collaboration for Improving Explainability and Reliability of Feedback in Introductory Programming Classrooms
& P5--E10, P1--E2, P1--E6, P9--E3, P9--E10
& \cite{hoq_explainable_nodate} \\

Explaining Code with a Purpose: An Integrated Approach for Developing Code Comprehension and Prompting Skills
& P6--E1, P6--E4, P11--E5
& \cite{denny_explaining_2024} \\ 

Exploration of Computer Programming Teaching Reform Based on Large Language Models
& P5--E4, P9--E3, P9--E7, P9--E4, P9--E9, P9--E10
& \cite{zhao_exploration_2025} \\ 

Exploring GenAI as a Tutoring Tool: A Case Study in First-Year Computer Programming
& P6--E1, P9--E1, P2--E1, P3--E1, P4--E7
& \cite{stienstra_exploring_2025} \\ 

Exploring How Multiple Levels of GPT-Generated Programming Hints Support or Disappoint Novices
& P3--E2, P3--E1, P3--E7, P2--E1, P9--E8
& \cite{xiao_exploring_2024} \\ 

Exploring the Effect of Scaffolding Strategies in GenAI Chatbot on Student Engagement and Programming Skill Development
& P3--E1, P1--E2, P10--E1, P4--E2, P1--E1, P6--E1
& \cite{khor_exploring_2025} \\

FastFixer: An Efficient and Effective Approach for Repairing Programming Assignments
& P9--E1, P5--E4, P3--E2
& \cite{liu_fastfixer_2024} \\ 

FEEDBOT: Formative Design Feedback on Programming Assignments
& P3--E1, P3--E6, P1--E7, P4--E6, P5--E1
& \cite{zhu_feedbot_2025} \\ 

Generative AI-Enabled Conversational Interaction to Support Self-Directed Learning Experiences in Transversal Computational Thinking
& P5--E1, P2--E1, P3--E1, P6--E1
& \cite{ouaazki_generative_2024} \\ 

Generative-AI Assisted Feedback Provisioning for Project-Based Learning in CS Courses
& P5--E1, P5--E9, P5--E2, P1--E3
& \cite{kusam_generative-ai_2024} \\ 

Improving AI in CS50: Leveraging Human Feedback for Better Learning
& P2--E1, P1--E3, P10--E1, P2--E2, P5--E10, P1--E2
& \cite{liu_improving_2025} \\ 

Improving the Efficacy of an Automated Judging System in an Introductory Programming Class using a Large Language Model
& P9--E2, P6--E5, P9--E3, P12--E3, P2--E1, P3--E1, P12--E1
& \cite{meza_improving_2024} \\ 

Improving User Engagement and Learning Outcomes in LLM-Based Python Tutor: A Study of PACE
& P3--E2, P9--E2, P9--E9, P9--E1, P1--E7, P3--E1
& \cite{shochcho_improving_2025} \\ 

Integrating AI Virtual Tutors in Blended Learning: A Controlled Evaluation of Learning Outcomes and Student Engagement
& P3--E1, P2--E1, P6--E1, P2--E2, P2--E10
& \cite{li_integrating_2025} \\ 

Intelligent Feedback for Individualized Introductory Programming Exercises
& P9--E1, P9--E4, P9--E7, P2--E6, P4--E11
& \cite{zampirolli_intelligent_2026} \\ 

Investigating the Impact of Codio Coach: A Specialized AI Learning Assistant on Computing Student Engagement and Performance
& P2--E1, P4--E1, P3--E1, P5--E1, P9--E7
& \cite{chandarana_investigating_2025} \\ 

It’s Dangerous to Prompt Alone! Exploring How Fine-Tuning GPT-4o Affects Novices’ Programming Error Resolution
& P2--E1, P5--E2, P10--E1
& \cite{santos_its_2025} \\ 

KOGI: A Seamless Integration of ChatGPT into Jupyter Environments for Programming Education
& P10--E1, P9--E1, P9--E4
& \cite{kuramitsu_kogi_2023} \\ 

Enhancing Algorithm Comprehension for Visually Impaired Individuals With LLM-Based Programming Code Segmentation
& P6--E1, P6--E8, P2--E6, P4--E5, P3--E1
& \cite{takamatsu_enhancing_2025} \\ 

Less stress, better scores, same learning: The dissociation of performance and learning in AI-supported programming education
& P3--E1, P3--E6, P2--E1, P1--E2
& \cite{bassner_less_2026} \\ 

Leveraging ChatGPT for Adaptive Learning through Personalized Prompt-based Instruction: A CS1 Education Case Study
& P12--E1, P1--E1, P2--E6, P11--E10, P9--E1, P10--E1
& \cite{abolnejadian_leveraging_2024} \\ 

Leveraging Large Language Models to Enhance Self-Regulated Learning in Programming Education With Explainable AI
& P5--E1, P9--E1, P10--E1
& \cite{yang_leveraging_2026} \\ 

LLM-based Interactive Coding Education via Predictive Query Management and Student-Centered Fine-Tuning: Design and Implementation with 1500-Student Class Data
& P9--E2, P5--E2, P12--E9, P9--E7, P1--E2
& \cite{youn_llm-based_2025} \\ 

LLM-Enhanced Learning Environments for CS: Exploring Data Structures and Algorithms with Gurukul
& P1--E3, P2--E1, P11--E1, P3--E6, P5--E1, P9--E3
& \cite{rachha_llm-enhanced_2024} \\ 

Navigating Compiler Errors with AI Assistance - A Study of GPT Hints in an Introductory Programming Course
& P3--E1, P3--E6, P3--E8, P9--E1, P4--E4
& \cite{pankiewicz_navigating_2024} \\ 

Next-Step Hint Generation for Introductory Programming Using Large Language Models
& P3--E1, P10--E1, P1--E2, P9--E9
& \cite{roest_next-step_2024} \\ 

Not the Silver Bullet: LLM-enhanced Programming Error Messages are Ineffective in Practice
& P5--E1, P5--E7
& \cite{santos_not_2024} \\ 

OI-Assistant: A Retrieval Augmented System for Similar Problem Discovery and Interactive Learning in Competitive Programming
& P12--E3, P5--E4, P9--E3, P3--E1, P3--E8, P3--E4
& \cite{su_oi-assistant_2025} \\ 

On Teaching Novices Computational Thinking by Utilizing Large Language Models Within Assessments
& P5--E1, P2--E1, P3--E1
& \cite{hassan_teaching_2025} \\ 

One Step at a Time: Combining LLMs and Static Analysis to Generate Next-Step Hints for Programming Tasks
& P3--E1, P3--E6, P1--E1
& \cite{birillo_one_2024} \\ 

Owlgorithm: Supporting Self-Regulated Learning in Competitive Programming through LLM-Driven Reflection
& P9--E7, P6--E1, P12--E1, P2--E1, P1--E9, P6--E7
& \cite{nieto-cardenas_owlgorithm_2026} \\ 

Pacing for Mastery: Optimizing LLM Interactions for Learning
& P1--E1, P9--E7, P2--E1, P2--E10, P3--E1
& \cite{tran_pacing_2026} \\ 

PersonaGPT: A Context-Aware Personalization Engine for Educational Chatbots Using Dynamic Learner Personas and Reflexive Dialogue
& P10--E1, P9--E2, P9--E1, P10--E2, P3--E1, P3--E7
& \cite{troussas_personagpt_2025} \\ 

Playing Dumb to Get Smart: Creating and Evaluating an LLM-based Teachable Agent within University Computer Science Classes
& P5--E1, P1--E1, P6--E1
& \cite{rogers_playing_2025} \\ 

Prompt Programming: A Platform for Dialogue-based Computational Problem Solving with Generative AI Models
& P1--E1, P6--E6, P6--E1, P2--E1, P11--E6
& \cite{padurean_prompt_2025} \\ 

Prompting for Engagement: Using the ICAP Framework to Guide Prompt Design in LLM-Powered Dialogue-Based Tutoring System for Novice Programmers
& P5--E1, P3--E1, P1--E1
& \cite{perez_prompting_2025} \\ 

PromptTutor: Effects of an LLM-Based Chatbot on Learning Outcomes and Motivation in Flipped Classrooms
& P1--E1, P5--E1, P6--E1, P9--E1, P6--E4, P3--E1, P6--E11
& \cite{zhang_prompttutor_2025} \\ 

Quack the Code: A Computer Game Show Offers Learning Through Teaching AI in Undergraduate Software Engineering
& P10--E8, P9--E8, P9--E7, P12--E1, P1--E1, P8--E8, P8--E6
& \cite{kelly_quack_2025} \\ 

ReDefining Code Comprehension: Function Naming as a Mechanism for Evaluating Code Comprehension
& P5--E6, P1--E2, P3--E4
& \cite{smith_redefining_2025} \\ 

Scaffolding Computational Thinking With ChatGPT
& P3--E2, P1--E10, P3--E5, P4--E4
& \cite{liao_scaffolding_2024} \\ 

Supporting Peer-to-Peer Learning with LLMs: Investigating Smarter Student Solution Recommendations
& P9--E1, P3--E1, P4--E1, P7--E3, P6--E1, P10--E1, P7--E1
& \cite{wiktor_supporting_2026} \\ 

Supporting Students in Prototyping AI-backed Software with Hosted Prompt Template APIs
& P11--E8, P6--E9, P7--E8
& \cite{aveni_supporting_2025} \\ 

Tailoring Your Code Companion: Leveraging LLMs and RAG to Develop a Chatbot to Support Students in a Programming Course
& P9--E1
& \cite{alario-hoyos_tailoring_2024} \\ 

Teach AI How to Code: Using Large Language Models as Teachable Agents for Programming Education
& P1--E1, P6--E7, P6--E2, P9--E7
& \cite{jin_teach_2024} \\ 

Teaching CS50 with AI: Leveraging Generative Artificial Intelligence in Computer Science Education
& P2--E1, P1--E1, P1--E2, P1--E10, P9--E1
& \cite{liu_teaching_2024} \\ 

That's Not the Feedback I Need! - Student Engagement with GenAI Feedback in the Tutor Kai
& P4--E6, P5--E1, P3--E1
& \cite{jacobs_thats_2025} \\ 

The Power of Context: An LLM-based Programming Tutor with Focused and Proactive Feedback
& P1--E1, P2--E1, P9--E7, P9--E9
& \cite{mueller_power_2025} \\ 

Towards Automated Interactive Tutoring - Focussing on Misconceptions and Adaptive Level-Specific Feedback
& P5--E4, P5--E5, P5--E1, P3--E6, P9--E6
& \cite{jell_towards_2023} \\ 

TraceMate: Collaborating with AI in Test-Driven Programming
& P3--E1, P3--E4
& \cite{wu_tracemate_2025} \\ 

Training AI Model that Suggests Python Code from Student Requests in Natural Language
& P6--E1, P1--E2, P6--E2, P10--E1, P3--E1, P2--E2
& \cite{kuramitsu_training_2024} \\ 

Unlimited Practice Opportunities: Automated Generation of Comprehensive, Personalized Programming Tasks
& P12--E1, P9--E1, P4--E11, P1--E5, P2--E6
& \cite{jacobs_unlimited_2025} \\ 

Unlocking Learning Potential: Generative AI Chatbots as Study Partners in Online BS in Computer Science Degree Program
& P1--E3, P5--E3
& \cite{qu_unlocking_2024} \\ 

UnrealMentor GPT: A System for Teaching Programming Based on a Large Language Model
& P5--E2, P9--E1, P9--E9, P5--E3, P9--E2, P9--E7
& \cite{zhu_unrealmentor_2025} \\ 

Web-based programming challenge platform using generative AI to improve programming skills in undergraduate engineering students
& P12--E2, P9--E2, P8--E8, P9--E1, P9--E8, P9--E4, P7--E8
& \cite{yarleque_web-based_2025} \\ 

What You Need is what You Get: Theory of Mind for an LLM-Based Code Understanding Assistant
& P9--E1
& \cite{richards_what_2024} \\ 

\bottomrule
\end{tabular}%
}
\end{table*}

\vspace{12pt}

\vspace{12pt}

\end{document}